\documentclass{jfm}
\usepackage{setspace}
\usepackage{graphicx}
\usepackage{subcaption}
\usepackage{overpic}
\usepackage{newtxtext}
\usepackage{newtxmath}
\usepackage{natbib}
\usepackage{arydshln}
\usepackage{color}
\usepackage{hyperref}
\hypersetup{
    colorlinks = true,
    urlcolor   = blue,
    citecolor  = black,
}

\usepackage{cleveref}
\crefname{section}{§}{§§}

\newcommand{\RomanNumeralCaps}[1]
\title{Biglobal linear stability analysis of a ducted 2D premixed flame: intrinsic thermoacoustic mode and role of exceptional point}

\author{Lu Chen\aff{1,}\aff{2}
 \and Yu Lv\aff{1,}\aff{3}
 \corresp{\email{lvyu@imech.ac.cn}}}

\affiliation{
\aff{1}State Key Laboratory of Nonlinear Mechanics, Institute of Mechanics, Chinese Academy of Sciences, Beijing, 100190, China
\aff{2}Department of Mechanical Engineering, National University of Singapore, 9 Engineering Drive 1,
117575, Republic of Singapore
\aff{3}School of Engineering Sciences, University of Chinese Academy of Sciences, Beijing, 101408, China
}

\begin{document}
\maketitle

\begin{abstract}
In this work, we aim to establish a detailed description of the physical mechanisms of intrinsic thermoacoustic modes and their interplay with duct acoustic modes. Tsisbiglobal linear stability  of an anchored laminar flame in an acoustic duct is carried out by exploiting the linearized compressible reactive flow equations. The pure intrinsic thermoacoustic mode is first identified and characterized under conditions where the eigenfrequencies of duct acoustic modes are sufficiently high and no mode interplay is present. Parameter variations, considering the inflow Mach number and acoustic reflection coefficients, are conducted to study the behaviors of eigenmode trajectories in the proximity of an exceptional point. Near the exceptional point, trajectory veering and mode switching are observed, and the extreme sensitivity renders the mode trajectory highly dependent on the choices of parameter and parameter variation path. One important finding is that different parameter variation paths can lead to inconsistent results in mode origin identification. Hence, we propose to characterize the thermoacoustic modes based on the flow structures. Specifically, the Helmholtz decomposition is employed to extract the potential and solenoidal components of the thermoacoustic modes. Intrinsic thermoacoustic modes and duct acoustic modes exhibit distinct flow structures, which are clearly distinguishable through the decomposed fields. Near the exceptional point, thermoacoustic modes display characteristics common to both intrinsic and duct modes. The results from compressible reactive flow analysis reveal that the flow structure of eigenmodes offers new insights into intrinsic thermoacoustic modes and mode interplay near the exceptional point.
\end{abstract}


% for manuscript review
%\doublespacing

\section{Introduction}
Thermoacoustic instability significantly impacts the design and operation of combustion systems, arising from the complex interplay between unsteady heat release and acoustic waves. This phenomenon is characterized by the perturbation of the flame caused by acoustic waves, which in turn induces heat release fluctuations. These fluctuations act as source terms for the acoustic field, thereby establishing a feedback mechanism. Within this loop, the energy released by the flame is converted into acoustic wave energy. The combustion system becomes susceptible to instability when the acoustic energy is insufficiently dissipated, leading to a sustained or growing oscillatory state. This instability phenomenon was observed and first analyzed one century ago~\citep{Rayleigh1878}. Due to high nonlinearity and sensitivity of the flame-acoustic interaction behaviors~\citep{juniper2018sensitivity}, thermoacoustic instability often exhibits rather complicated modality. Thereby, up to today, the prediction and control of thermoacoustic instability remains one of the most severe problems in designing modern propulsion systems~\citep{lieuwen2005combustion,poinsot2017prediction}.

It has long been considered that an acoustically reflective boundary is essential for the interaction of acoustic waves and flames within a combustion chamber, thereby giving rise to thermoacoustic modes. These eigenmodes are traditionally linked to the natural acoustic eigenmodes of the combustion chamber, often referred to as duct acoustic (AC) modes or thermoacoustic modes born of AC modes. However, recent advances in our understanding of thermoacoustic instability, particularly the discovery of intrinsic thermoacoustic (ITA) instability~\citep{hoeijmakers2014intrinsic}, have challenged this view~\citep{schuller2020dynamics,silva2023intrinsic}. ITA instability manifests in an anechoic configuration, where thermoacoustic modes are generated without the need for reflective boundary conditions. In this scenario, upstream acoustic waves produced by the flame perturb the velocity field ahead of the flame, leading to fluctuations in heat release rate. This establishes a feedback loop independent of boundary reflections. Intrinsic thermoacoustic instability is observed not only in purely anechoic environments, where the eigenmode is termed a pure ITA mode, but also in combustion systems with reflective boundaries, where it is referred to as an ITA mode or thermoacoustic mode arising from an ITA mode. In general, the total number of thermoacoustic modes in a system is the sum of AC modes and ITA modes~\citep{emmert2017acoustic}.

The AC and ITA modes are governed by distinct physical mechanisms. AC modes, which can be regarded as the natural acoustic modes of the combustor, are modulated by the unsteady heat release from the flame. The geometric boundary conditions at the inlet and outlet are critical in determining the characteristics of AC modes. In contrast, ITA modes are intrinsically linked to the flame dynamics, particularly the flame's response to upstream velocity perturbations and the consequent generation of heat fluctuations. The hydrodynamic flows near the flame, along with fuel properties, significantly influence the characteristics of ITA modes~\citep{dupuy2024control}. Given the independence of these two distinct thermoacoustic modes, their characteristic frequencies may either be closely spaced or widely separated, depending on the boundary conditions and hydrodynamic flow characteristics. This suggests that in the eigenspectrum of thermoacoustic systems, branches of AC and ITA eigenmodes, each with certain parametric dependencies, can potentially coalesce at a specific point. This point, known as the exceptional point, marks the location where the thermoacoustic system becomes non-Hermitian. Mathematically, the exceptional point is defined as a singularity in the eigenspectrum of linear operators, where the system's sensitivity to parameters becomes infinite~\citep{kato2013perturbation}. The concept of the exceptional point has garnered significant attention in physics~\citep{heiss2012physics}, and its presence in thermoacoustic systems has been reported and analyzed over the recent years~\citep{mensah2018exceptional, orchini2020thermoacoustic,magri2023linear}. Interestingly, the exceptional point has even been utilized to design the stable thermoacoustic system and choose proper operating parameters~\citep{casel2024novela, casel2024novelb}. The extreme parameter sensitivity near the exceptional point can impact the accuracy and reliability of numerical stability analysis~\citep{schaefer2021impact}. Overall,  the exceptional point plays a rather important role in identifying mode origins and decoding the behaviors of mode interplay. 

In previous studies on exceptional points in thermoacoustic systems, the acoustic network framework has typically been employed~\citep{sogaro2019thermoacoustic, ghani2021exceptional}. Within this approach, acoustic fields are described using the Helmholtz equations, and occasionally, the linearized Euler equations~\citep{aguilar2017adjoint}. The flame is often modeled by a flame transfer function that uses only the streamwise velocity at a reference point as input. To obtain the corresponding thermoacoustic eigenmodes, a nonlinear eigenvalue problem must be solved~\citep{nicoud2007acoustic}. While this method is practical for many applications, it may be excessively simplified in analyzing the physical nature of ITA mode and its interplay with AC mode, because the flame is represented by a scalar function, decoupled from the hydrodynamic and acoustic fields. Such simplification may neglect critical physical aspects which may be essential to the presence of ITA mode. One improvement to this limitation involves coupling the reactive flow equations in the low Mach number limit with the Helmholtz equations to describe the acoustic fields~\citep{magri2014global, magri2017multiple}. However, this approach still encounters challenges in defining the appropriate coupling relations between the reactive flow fields and the acoustic fields. More recently, the compressible Navier-Stokes equations, combined with flame models and biglobal stability analysis, have been employed to investigate the stability properties of premixed flames~\citep{wang2022linear, varillon2023global, brokof2024role}. These studies discuss the influences of hydrodynamic flows on intrinsic thermoacoustic instability, yet the concept of exceptional points and the interaction between thermoacoustic modes arising from ITA modes and AC modes are rarely addressed within the biglobal stability analysis framework. High-order sensitivities and the convergence ranges of eigenvalue expansions concerning parameters present another method for analyzing exceptional points and mode interplay. However, this approach remains limited to the Helmholtz equations~\citep{orchini2020degenerate} or the incompressible Navier-Stokes equations~\citep{knechtel2024arbitrary}. To the best of the authors’ knowledge, no existing literature addresses exceptional points or mode switching between different types of thermoacoustic modes within the framework of compressible reactive flows. By employing this method, we would like to gain deeper insights into the flow and acoustic fields, rather than simplifying the flame to mere jump conditions in acoustic networks.

In the analysis of the thermoacoustic instability, parameter variation has often been employed to determine whether an unstable thermoacoustic mode originates from an ITA mode or an AC mode~\citep{emmert2017acoustic, mukherjee2017intrinsic, hosseini2018intrinsic}. This technique has revealed that the trajectories of eigenmodes can veer and even undergo mode switching near the exceptional point~\citep{sogaro2019thermoacoustic}. However, the high parameter sensitivity in the vicinity of the exceptional point often leads to inconsistent identification of thermoacoustic modes, depending on the choice of parameters~\citep{silva2023intrinsic}. Given these limitations, a phasor diagram based method has been proposed to discriminate different types of thermoacoustic modes with their mode shapes~\citep{yong2021categorization, yong2023categorization}. This approach has been applied to ideal combustor and flame models represented by acoustic network models, which inspires us to foucs on the characteristic mode structures of thermoacoustic modes.

In the present study, we investigate thermoacoustic modes, focusing on characterizing mode origins and the role of exceptional points in mode interplay, within the framework of linearized compressible Navier-Stokes equations. This comprehensive framework offers a unique capability to examine critical features of both acoustic and hydrodynamic fields, enabling their association with ITA and AC eigenmodes. Mode identification will be conducted with the aid of parameter variations, with consideration given to different paths of variation. To elucidate the inherent physical characteristics of each eigenmode, we investigate its mode shape and propose applying Helmholtz decomposition to post-process corresponding mode structure. Helmholtz decomposition has been employed in studies of compressible isotropic turbulence~\citep{samtaney2001direct} and wall turbulence~\citep{yu2019genuine} to isolate the contributions of acoustic and vortical components to turbulence dynamics. Recently, this decomposition method has been utilized in the post-processing of aeroacoustic computations to decouple acoustic fields from hydrodynamic flows~\citep{schoder2019hybrid} or investigate the influence of thermal expansion on the unburned mixture in a turbulent premixed flame~\citep{sabelnikov2021application}. A similar approach, based on isolating the potential and rotational components of velocity fields, has been applied to analyze the response of flames to velocity perturbations~\citep{STEINBACHER20195367}, but it has not yet been employed to analyze thermoacoustic eigenvalue problems. Here, we propose utilizing this decomposition approach to extract potential and solenoidal features from thermoacoustic eigenmodes, with the goal of advancing our understanding of mode switching phenomena and the influence of exceptional points.

The paper is structured as follows: In~\cref{Problem formulation}, we formulate the problem by introducing the flame configuration, governing equations, boundary conditions, and the methods of linear analysis, including the Helmholtz decomposition. Numerical details relevant to the study are also provided in this section. \cref{Intrinsic thermoacoustic mode} presents an investigation of ITA modes under both anechoic and fully reflective boundary conditions, laying the groundwork for subsequent research on the interplay between different types of thermoacoustic eigenmodes. \cref{identification exceptional points} focuses on the identification of exceptional points in the thermoacoustic eigenspectra through parameter variation, concerning the challenges associated with using parameter variation to classify thermoacoustic modes. In~\cref{Mode structure}, we investigate characteristic mode structures of thermoacoustic modes of different origins. Besides, \cref{Mode structure} includes the results of Helmholtz decomposition of various thermoacoustic eigenmodes, highlighting their distinct physical characteristics. Finally, the conclusions are summarized in~\cref{Conclusions}

\section{Mathematical Description}
\label{Problem formulation}
\subsection{Flame configuration}
\begin{figure}
    \begin{subfigure}{0.9\textwidth}
        \centering
        \centerline{
        \begin{overpic}[width=1\textwidth]{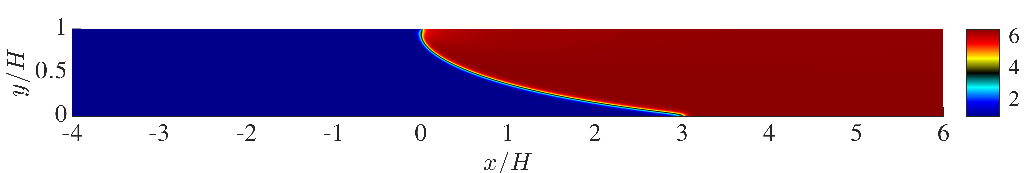}
        \put(2.5,16){(\textit{a})}
        \end{overpic}
        }
        \phantomsubcaption        
    \end{subfigure}
    \vspace{0pt}
    \begin{subfigure}{0.9\textwidth}
        \centering
        \centerline{
        \begin{overpic}[width=1\textwidth]{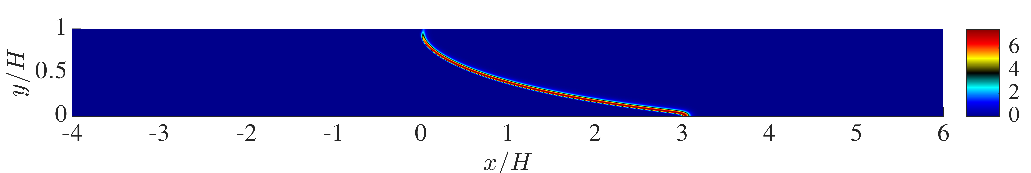}
        \put(2.5,16){(\textit{b})}
        \end{overpic}
        }
        \phantomsubcaption        
    \end{subfigure}   
\caption{Baseflow solution of the flame in an acoustic duct: 
  (\textit{a}) temperature profile and (\textit{b}) heat release rate profile.}
\label{fig:baseflow}
\end{figure}

We investigate a 2D laminar premixed flame anchored in an acoustic duct, as depicted in figure~\ref{fig:baseflow}. The duct has a length $L$ and a half-height $H$. The top wall is divided into a cold wall and a hot wall at $x=0$. In our study, the duct length $L$ is varied, and two different configurations are considered: $L/H=10$ and $L/H=50$. In the first configuration ($L/H=10$), the frequency of the first AC mode is sufficiently high, lying above the frequency range of the ITA mode. Our focus in this setting will be on the properties of the ITA mode. In the second configuration ($L/H=50$), as the length of duct increases, the frequency of the AC mode approaches that of the ITA mode, allowing us to investigate the interplay between these two types of thermoacoustic modes. The steady state flame profile of the first configuration is illustrated in figure~\ref{fig:baseflow}, which is similar to previous studies on 2D anchored flames~\citep{meindl2021spurious}.

\subsection{Governing equations}
The flame characteristics are governing by the physics of chemically reactive flows. As such, the compressible Navier-Stokes equations with global one-step chemical mechanism are employed to investigate the thermoacoustic system of a laminar premixed flame anchored at an acoustic duct. The governing equations read: 

\begin{equation}
	\frac{\partial \rho}{\partial t}+\bnabla\bcdot (\rho\boldsymbol{u})=0 \;,
\end{equation}
\begin{equation}
	\frac{\partial (\rho \boldsymbol{u})}{\partial t}+\bnabla\bcdot (\rho\boldsymbol{u}\boldsymbol{u})=-\bnabla p + \bnabla \cdot \boldsymbol{\tau}  \;, 
\end{equation}
\begin{equation}
	\frac{\partial (\rho Y_{F})}{\partial t}+\bnabla\bcdot (\rho\boldsymbol{u}Y_{F})= \bnabla \cdot (\rho D \bnabla Y_{F}) + \dot{\omega}_{F} \;, 
\end{equation}
\begin{equation}
	\frac{\partial (\rho h_{s})}{\partial t}+\bnabla\bcdot (\rho\boldsymbol{u}h_{s})= \frac{\partial p}{\partial t} + \boldsymbol{u}\cdot\bnabla{p} + \bnabla \cdot (\lambda \bnabla T) + \boldsymbol{\tau}:\bnabla\boldsymbol{u} + \dot{\omega}_{T} \;.
\end{equation}
in which $\rho$ is the density, $\boldsymbol{u}=(u_{x},u_{y})$ is the velocity vector with the streamwise ($u_x$) and normal ($u_y$) components, $p$ is the pressure, $Y_{F}$ is the mass fraction of fuel and $h_{s}$ stands for the sensible enthalpy. The specific heat capacity $C_{p}$ is assumed to be constant for all species as $C_{p} = 1.004~\mathrm{kJ/(kg\cdot K)}$. The sensible enthalpy is expressed as $h_{s} = C_{p}T$, where $T$ denotes the temperature. The viscous stress tensor is written as:
\begin{equation}
  \boldsymbol{\tau} = \mu(\bnabla\boldsymbol{u}+\bnabla\boldsymbol{u}^{T}-\frac{2}{3}(\bnabla\bcdot\boldsymbol{u})\boldsymbol{I}) \;,
\end{equation}
where $\mu$ is the viscosity evaluated using Sutherland's law, $\mu = A_{s} T^{1/2}/(1+T_{s}/T)$ with $A_{s} = 1.7 \times 10^{-6}~\mathrm{kg/(m \cdot s \cdot K^{1/2})}$ and $T_{s}=170.7$ K. The thermal conductivity $\lambda$ and species mass diffusivity $D$ are determined through the Prandtl number $Pr = \mu C_{p}/ \lambda $ and Lewis number $ Le = \lambda / (C_{p}\rho D)$. The Prandt number and Lewis numbers are set to be constants in this study, $Pr = 0.7$ and $ Le = 1.0$. The one-step global reaction scheme 1S$_{-}$CH4$_{-}$MP1~\citep{1SCH4MP1} is selected and the reaction rate $\dot\omega$ is modeled using the Arrhenius law:
\begin{equation}
\dot\omega = A_{r}[X_{F}]^{n_{F}}[X_{O}]^{n_{O}}\mathrm{exp}(-\frac{T_{a}}{T}) \;,
\end{equation}
Here Arrhenius pre-exponential factor $A_{r} = 1.1\times 10^{7}~\mathrm{m^{3/2}/(s\cdot mol^{1/2})}$ and activation temperature $T_{a} = 1.0065\times 10^{4}$ K. The exponent numbers for reactants are $n_{F}=1$ and $n_{O}=0.5$. The molecular masses are given by $W_{F}=16$ g/mol and $W_{O}=32$ g/mol. The concentrations are evalated using the mass fractions, $[X_{F}]=\rho Y_{F}/W_{F}$ and $[X_{O}]=\rho Y_{O}/W_{O}$. A lean combustible mixture is considered, and the mass fraction of oxidizer $Y_{O}$ can be evaluated with the equivalence ratio $\phi$ at the inlet boundary and stoichiometric ratio $s = 4.0$ as: $Y_{O}= s(Y_{F}+1/\phi-1)$. The source terms in the specis equation and energy equation are expressed as $\dot\omega_{F}=-W_{F}\dot\omega$ and $\dot\omega_{T}=-\Delta h_{f}^{o} \dot\omega$ with the reaction enthalpy $\Delta h_{f}^{o}=-804$~kJ/mol. This chemical scheme for methane-air reaction has been employed in the recent studies of flame instability~\citep{wang2022global,wang2024onset}.

The flow variables are non-dimensionalized with the inflow streamwise velocity $u_{0}$, the half height of the acoustic duct $H$, the inflow density $\rho_{0}$, the inflow mass fraction of fuel $Y_{F,0}$, the inflow temperature $T_{0}$, the ambient pressure $p_{0}$ and the viscosity of the unburnt mixture $\mu_{0}$. The dimensionless pressure $\widetilde{p}$ is expressed as: $\widetilde{p}=(p-p_{0})/(\rho_{0}u_{0}^{2})$ and the notation $\: \widetilde{} \:$ will be omitted in the following text. The Reynolds number is defined as $Re = (\rho_{0}u_{0}H)/\mu_{0}$. The adiabatic temperature change $\Delta T$ is evaluated as $\Delta T=(Y_{F,0}\Delta h_{f}^{o})/(W_{F}C_{p}T_{0})$ and the Zeldovich number is defined as $Ze=T_{a}\Delta T/(T_{0}(1+\Delta T)^{2})$. The Damkohler number is expressed as $Da=(A_{r}H/u_{0})\sqrt{(\rho_{0}Y_{F,0})/W_{O}}$. The Mach number is defined as $Ma = \sqrt{(\rho_{0}u_{0}^{2})/(\gamma p_{0})}$, where $\gamma$ denotes the specific heat ratio and is set to a constant $\gamma = 1.4$. In our study, we only adjust the Mach number while fixing other non-dimensional parameters to be: $Re=500$, $Ze=4.35$ and $Da=1.7\times 10^{5}$. This set of parameters are selected to qualitatively represent a laminar premixed flame, rather than matching any specific experimental condition, similar to the previous researches on premixed conical flames~\citep{douglas2023flash}. However, the non-dimensional parameters may be deduced from a experimental or computational setup to facilitate such a combustion analysis. 

\subsection{Linear analysis}
In order to introduce the linearization for the reactive flow system, the governing equations are formulated into:
\begin{equation}
\boldsymbol{M}\frac{\partial \boldsymbol{q}}{\partial t} = \boldsymbol{N}(\boldsymbol{q}) \;,
\end{equation}
where $\boldsymbol{q}$ denotes the flow variables, $\boldsymbol{q} = (u_{x},u_{y},Y_{F},T,p)^{T}$. The flow variables are decomposed into the steady part $\overline{\boldsymbol{q}}$ and fluctuating part $\boldsymbol{q}'$: 
\begin{equation}
\boldsymbol{q}=\overline{\boldsymbol{q}}+\boldsymbol{q}' \;, 
\end{equation}
The steady solution refers to the baseflow solution $\overline{\boldsymbol{q}}$, satisfying:
\begin{equation}
\boldsymbol{N}(\overline{\boldsymbol{q}}) = 0 \;, 
\end{equation}
The governing equations are linearized with respect to $\overline{\boldsymbol{q}}$:
\begin{equation}
\label{eqn_linearized_sys}
\boldsymbol{M}\frac{\partial \boldsymbol{q}'}{\partial t} = \boldsymbol{N}_{\overline{\boldsymbol{q}}}\boldsymbol{q}'\;,
\end{equation}
The fluctuation part is expanded with the eigenvector basis $\boldsymbol{q}'(x,y,t) = \hat{\boldsymbol{q}}(x,y)$exp$((\sigma+i\omega)t)$, where $\sigma$ denotes the growth rate and $\omega$ is the frequency. The non-dimensional Strouhal number is defined as $St = \omega H/u_{0}$. As such, Equation~(\ref{eqn_linearized_sys}) is expressed as a generalised eigenvalue problem:
\begin{equation}
(\sigma+i\omega)\boldsymbol{M}\hat{\boldsymbol{q}} = \boldsymbol{N}_{\overline{\boldsymbol{q}}}\hat{\boldsymbol{q}}\;,
\end{equation}
of which the solution constitutes the eigenmodes of the thermoacoustic system. The eigenmode is linearly unstable when the growth rate $\sigma$ is positive; otherwise, the eigenmode is linearly stable.

Besides linear stability analysis, the linear input-output analysis is also employed to evaluate the flame transfer function. Here the input forcing of frequency $\omega$ is denoted as $\hat{\boldsymbol{F}}$ and the input-output relation is written as:
\begin{equation}
i\omega\boldsymbol{M}\hat{\boldsymbol{q}} = \boldsymbol{N}_{\overline{\boldsymbol{q}}}\hat{\boldsymbol{q}} + \hat{\boldsymbol{F}}\;, 
\end{equation}
The output response fields $\hat{\boldsymbol{q}}$ could obtained by solving this system with input forcing $\hat{\boldsymbol{F}}$:
\begin{equation}
\hat{\boldsymbol{q}} =(i\omega \boldsymbol{M} - \boldsymbol{N}_{\overline{\boldsymbol{q}}})^{-1}\hat{\boldsymbol{F}} \;.
\end{equation}
With this input-output relation, the flame transfer function $\mathcal{F}(\omega)$ can be evaluated as:
\begin{equation}
\mathcal{F}(\omega) = \frac{\hat{\dot{Q}}_{T}(\omega)/\overline{\dot{Q}}_{T}}{\hat{u}_{x}(\omega,\boldsymbol{x}_{\mathrm{ref}})/\overline{u}_{x}(\boldsymbol{x}_{\mathrm{ref}})} \;, 
\end{equation}
in which $\overline{\dot{Q}}_{T}$ and $\hat{\dot{Q}}_{T}(\omega)$ refer, respectively, to the steady and fluctuation parts of the global heat release rate over the whole domain. $\overline{u}_{x}(\boldsymbol{x}_{\mathrm{ref}})$ and $\hat{u}_{x}(\omega,~\boldsymbol{x}_{\mathrm{ref}})$ correspond to the steady and fluctuation parts of the streamwise velocity located at a reference point $\boldsymbol{x}_{\mathrm{ref}}$ close to and upstream from the flame front. It is set as $\boldsymbol{x}_{\mathrm{ref}}/H = (-0.025,0.5)$ in the present study. The velocity perturbations are imposed at the inlet boundary as the input forcing $\hat{\boldsymbol{F}}$ and other quantities such as the global heat release rate $\hat{\dot{Q}}_{T}(\omega)$ and reference velocity $\hat{u}_{x}(\omega,\boldsymbol{x}_{\mathrm{ref}})$ are measured from the output fields. It has been shown in previous studies~\citep{meindl2021spurious,wang2022linear} that flame transfer function constructed with linear input-output analysis agrees well with that determined from transient simulation with system identification method.

\subsection{Helmholtz decomposition}
Helmholtz decomposition is conducted to the velocity field $\hat{\boldsymbol{u}}$ of eigenmode $\hat{\boldsymbol{q}}$ :
\begin{equation}
\hat{\boldsymbol{u}} = \hat{\boldsymbol{u}}_{d} + \hat{\boldsymbol{u}}_{s} \;,
\end{equation}
Here $\hat{\boldsymbol{u}}_{d}$ refers to the potential(dilatational) component and $\hat{\boldsymbol{u}}_{s}$ denotes the solenoidal(vortical) component of the velocity field. They satisfy $\hat{\boldsymbol{u}}_{d} = \bnabla \hat{\varphi}$ and $\hat{\boldsymbol{u}}_{s} = \bnabla \times \hat{\boldsymbol{\psi}}$. Here $\hat{\varphi} $ is the velocity potential and $\hat{\boldsymbol{\psi}}$ is the vector potential, which could be obtained from the Poisson equation:
\begin{equation}
\bnabla^{2}\hat{\varphi} = \bnabla \cdot \hat{\boldsymbol{u}} \;, \quad \bnabla^{2} \hat{\boldsymbol{\psi}} = -\bnabla \times \hat{\boldsymbol{u}} \;.
\end{equation}
To ensure the decomposition is unique and resulting velocity fields are proper with enforced boundary conditions, the streamwise velocity of solenoidal component $\hat{u}_{s,x}$ is determined from the $\bnabla \times \hat{\boldsymbol{\psi}}$ and the potential part is substracted from the $\hat{u}_{d,x}=\hat{u}_{x}-\hat{u}_{s,x}$. While the normal velocity of potential part $\hat{u}_{d,y}$ is evaluated from the $\bnabla \hat{\varphi}$ and the solenoidal component $\hat{u}_{s,y}=\hat{u}_{y}-\hat{u}_{d,y}$. The boundary conditions for Helmholtz decomposition will be listed and discussed in Appendix~\ref{appA}.  

It is important to note that in the Helmholtz decomposition, all compressible effects are ascribed to the potential component. Beyond acoustic effects, significant thermal expansion near the flame front, as well as entropy and species-related perturbations, are not distinguished. This represents a limitation of the Helmholtz decomposition. Recently, \citet{brokof2023towards} introduced an improved Momentum Potential Theory to more precisely identify acoustic sources in flame analysis, addressing the shortcomings of Helmholtz decomposition. However, in our study, we do not seek to isolate individual contributions from various compressible effects. Instead, we focus on their collective influence on the stability of the thermoacoustic system. Therefore, Helmholtz decomposition remains a useful post-processing tool, allowing us to extract hydrodynamic effects into the solenoidal component while grouping all compressible effects into the potential part. In this context, Helmholtz decomposition serves primarily as a means to identify flow structures associated with different thermoacoustic eigenmodes.

\subsection{Additional solution details}

\begin{table}
  \begin{center}
\def~{\hphantom{0}}
  \begin{tabular}{lc}
      Boundary   & Constraints\\[3pt]
       Inlet   & $u_{x}=1.5(1-y^{2}),\ u_{y}=0,\ Y_{F}=1,\ T=1, \phi = 0.8$\\
       Cold Wall   & $u_{x}=u_{y}=0,\ \boldsymbol{n}\cdot\bnabla Y_{F}=0,\ T=1$\\
       Hot wall  & $u_{x}=u_{y}=0,\ \boldsymbol{n}\cdot\bnabla Y_{F}=\boldsymbol{n}\cdot\bnabla T=0$\\
       Symmetry Axis  & $\partial_{y}u_{x}=0, \ u_{y}=0,\ \boldsymbol{n}\cdot\bnabla Y_{F}=\boldsymbol{n}\cdot\bnabla T=0$\\
       Outlet & $(-p\boldsymbol{I}+\Rey^{-1}\bnabla\boldsymbol{u})\cdot\boldsymbol{n}=0,\ \boldsymbol{n}\cdot\bnabla Y_{F}=\boldsymbol{n}\cdot\bnabla T=0$\\
  \end{tabular}
  \caption{Boundary conditions for the baseflow calculation (recall that $\phi$ refers to the equivalence ratio at the inlet boundary).}
  \label{tab:boundary conditions for baseflow}
  \end{center}
\end{table}

\begin{table}
  \begin{center}
\def~{\hphantom{0}}
  \begin{tabular}{lc}
      Boundary   & Constraints\\[3pt]
       Inlet   & $\hat{f}_{in}-R_{in}\hat{g}_{in}=0,\ \hat{u}_{y}=0,\ \hat{Y}_{F}=0,\ \hat{T}=0$\\
       Cold Wall   & $\hat{u}_{x}=\hat{u}_{y}=0,\ \boldsymbol{n}\cdot\bnabla \hat{Y}_{F}= 0, \ \hat{T}=0$\\
       Hot wall  & $\hat{u}_{x}=\hat{u}_{y}=0,\ \boldsymbol{n}\cdot\bnabla \hat{Y}_{F}=\boldsymbol{n}\cdot\bnabla \hat{T}=0$\\
       Symmetry Axis  & $\partial_{y}\hat{u}_{x}=0, \ \hat{u}_{y}=0,\ \boldsymbol{n}\cdot\bnabla \hat{Y}_{F}=\boldsymbol{n}\cdot\bnabla \hat{T}=0$\\
       Outlet & $\hat{g}_{out}-R_{out}\hat{f}_{out}=0,\ \partial_{x}\hat{u}_{y} = 0, \ \boldsymbol{n}\cdot\bnabla \hat{Y}_{F}=\boldsymbol{n}\cdot\bnabla \hat{T}=0$\\
  \end{tabular}
  \caption{Boundary conditions for eigenvalue problem. }
  \label{tab:boundary conditions for evp}
  \end{center}
\end{table}

The boundary conditions for the calculation of baseflow $\overline{\boldsymbol{q}}$ are listed in table~\ref{tab:boundary conditions for baseflow}. The boundary conditions for the eigenvalue problems are listed in table~\ref{tab:boundary conditions for evp}. Here the acoustic characteristic waves $\hat{f}$ and $\hat{g}$ are introduced:
\begin{equation}
\hat{f} = \hat{p}+\mathop{\overline{\rho}}\mathop{\overline{c}}\hat{u}_{x}, \quad
\hat{g} = \hat{p}-\mathop{\overline{\rho}}\mathop{\overline{c}}\hat{u}_{x} \;, 
\end{equation}
in which $\overline{c}$ refers to the sound speed of the baseflow solutions. The boundary conditions at inlet and outlet are imposed with the reflection coefficients $R_{in}$ and $R_{out}$:
\begin{equation}
R_{in} =\frac{\hat{f}_{in}}{\hat{g}_{in}}, \quad R_{out}=\frac{\hat{g}_{out}}{\hat{f}_{out}} \; 
\end{equation}
where the subscript notation ``${in}$'' or ``$out$'' are introduced to denote the acoustic characteristic wave at the inlet or outlet boundary, respectively. At the outlet boundary, it is assumed that the perturbation level of normal velocity is in a much smaller magnitude, compared to the streamwise counterpart. In addition, the acoustic wave reflection introduced by the entropy wave at the outlet boundary is neglected, which will be discussed in details in Appendix~\ref{appB}.

The governing equations, combined with the boundary conditions, are discretized using the finite element method implemented in the open-source software FreeFem++~\citep{hecht2012new}, and all results are gathered using the open-source drivers StabFem~\citep{fabre2018practical}. Penalty terms are applied to the eigenvalue problem to enforce the boundary conditions for the acoustic characteristic waves, $\hat{f}$ and $\hat{g}$. All flow variables are discretized using P2 element except for pressure, which is approximated using P1 element. The velocity potential and vector potential in the Helmholtz decomposition are discretized using P3 element to ensure that the decomposed velocity fields are accurately represented by P2 elements. The computational mesh consists of approximately $10^{5}$ triangular elements, with mesh adaptation applied throughout the calculation. The flame front is resolved with at least 10 points. The baseflow solution is computed using Newton's method, and the eigenvalue problem is solved using the Krylov-Schur method, with a convergence tolerance set to $10^{-10}$. Numerical linear algebra tasks are performed using the open-source libraries PETSc~\citep{petsc-efficient} and SLEPc~\citep{Hernandez:2005:SSF}. The codes for calculating steady solutions and corresponding eigenmodes are validated in Appendix~\ref{appD}.

\section{Study on intrinsic thermoacoustic mode}
\label{Intrinsic thermoacoustic mode}
The objective of this section is to identify and analyze the ITA mode from the eigenspectrum of this duct flame thermoacoustic system. In particular, the key features of ITA mode in both anechoic and acoustically reflective duct configurations are examined and compared here.

\begin{figure}
  \centerline{\includegraphics[width=0.6\textwidth]{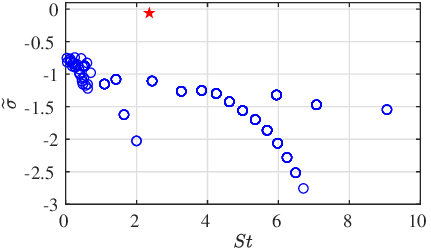}} 
  \caption{Eigenspectrum of the L10M1 case with $R_{in}=R_{out}=0$. The expected ITA mode is highlighted by a red pentagram. }
\label{fig:ev_L10_pureITA}
\end{figure}

\subsection{Pure intrinsic thermoacoustic mode}
\begin{figure}
  \centerline{\includegraphics[width=0.6\textwidth]{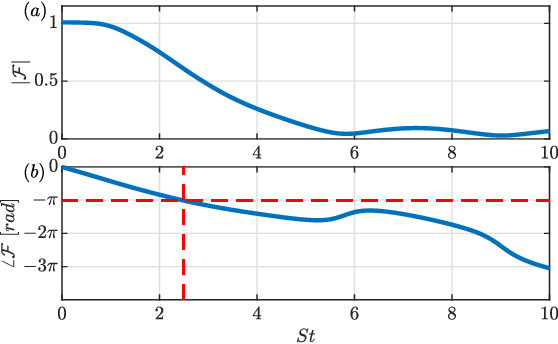}}
  \caption{Flame transfer function determined from the linearized input-output analysis.
  (\textit{a}) Gain and (\textit{b}) Phase with the blue solid lines. The frequency corresponding to $\angle \mathcal{F} = -\pi $ is indicated by the red dashed lines and is located at $St \approx 2.45$.}
\label{fig:FTF}
\end{figure}

We first investigate the anechoic configuration ($R_{in}=R_{out}=0$) with the duct length $L/H=10$ and an inflow Mach number $Ma$ = 0.01 (referred as case L10M1). The eigenspectrum is depicted in figure~\ref{fig:ev_L10_pureITA}, revealing that all eigenmodes are stable. The least stable eigenmode, with a frequency of  $St \approx 2.36$, is identified to be the pure ITA mode. This identification can be verified using the criterion established by the acoustic network models. In the acoustic network models, the frequency of the ITA mode in an anechoic environment is linked to the frequency at which the phase of the flame transfer function, $\mathcal{F}(\omega)$ equals $-\pi$~\citep{hoeijmakers2014intrinsic,hoeijmakers2016flame}. The gain and phase of $\mathcal{F}(\omega)$ obtained from linear input-output analysis are provided in figure~\ref{fig:FTF}. The eigenfrequency at $St \approx 2.36$ closely aligns with the frequency at which the phase lag of $-\pi$ occurs, thereby confirming that this eigenmode is indeed the ITA mode. In the following, we call this ITA mode the pure ITA mode as it exists in the anechoic duct configuration.

\begin{figure}
    \centering
    \begin{subfigure}[h]{0.82\textwidth}
        \begin{overpic}[width=1\textwidth]{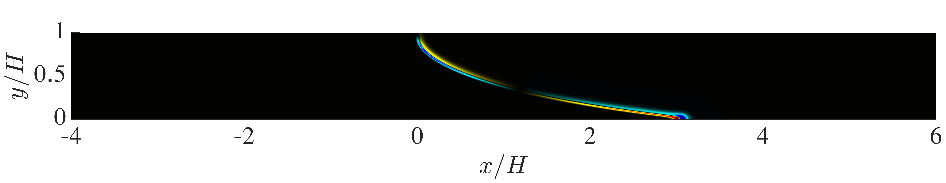}
        \put(2,17){(\textit{a})}
        \end{overpic}    
        \vspace{0pt}
        \begin{overpic}[width=1\textwidth]{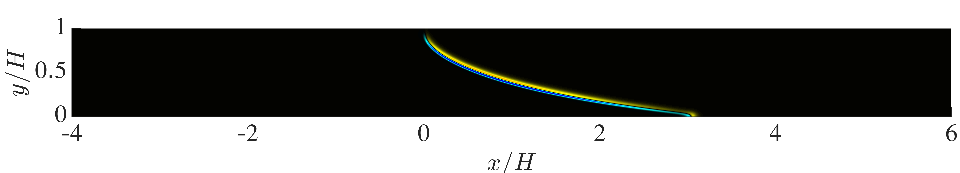}
        \put(2,17){(\textit{b})}
        \end{overpic}
        \phantomsubcaption
    \end{subfigure}%
    \hspace{5pt}                       
    \begin{subfigure}[h]{0.09\textwidth}
\includegraphics[width=\textwidth,height=0.225\textheight]{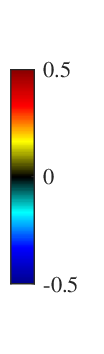}
    \phantomsubcaption
    \end{subfigure}  
    \caption{Heat release rate of pure ITA mode in case L10M1. (\textit{a}) Real part $\Real (\hat{\dot{\omega}}_{T})$  and (\textit{b}) Imaginary part $\Imag (\hat{\dot{\omega}}_{T}) $. All results are normalized by the maximum of $\left| \hat{\dot{\omega}}_{T} \right| $.}
\label{fig:pureITA_L10_Q}
\end{figure}

\begin{figure}
    \centering
    \begin{subfigure}[h]{0.82\textwidth}
        \begin{overpic}[width=1\textwidth]{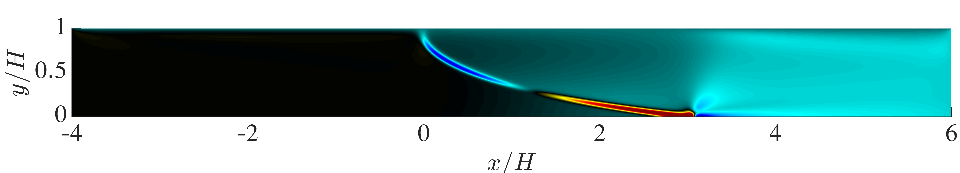}
        \put(2,17){(\textit{a})}
        \end{overpic}    
        \vspace{0pt}
        \begin{overpic}[width=1\textwidth]{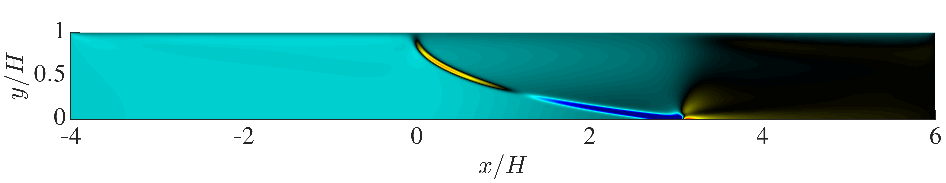}
        \put(2,17){(\textit{b})}
        \end{overpic}
        \phantomsubcaption
    \end{subfigure}%
    \hspace{5pt}                       
    \begin{subfigure}[h]{0.09\textwidth}
\includegraphics[width=\textwidth,height=0.225\textheight]{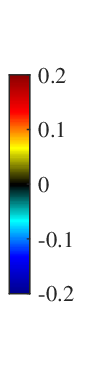}
    \phantomsubcaption
    \end{subfigure}  
\caption{Acoustic characteristic wave of pure ITA mode in case L10M1. Real part of forward acoustic characteristic wave $\hat{f}$ (\textit{a}). Real part of Backward acoustic characteristic wave $\hat{g}$ (\textit{b}). All results are normalized by its corresponding maximum of $\left| \hat{f} \right| $ or $\left| \hat{g} \right| $.}
\label{fig:pureITA_L10_ACwave}
\end{figure}

\begin{figure}
    \centering
    \begin{subfigure}[b]{0.82\textwidth}
        \begin{overpic}[width=1\textwidth]{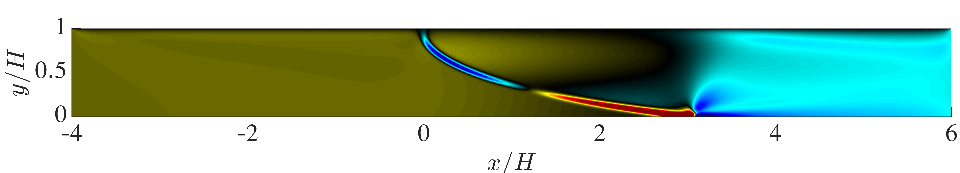}
        \put(2,17){(\textit{a})}
        \end{overpic}    
        \vspace{0pt}
        \begin{overpic}[width=1\textwidth]{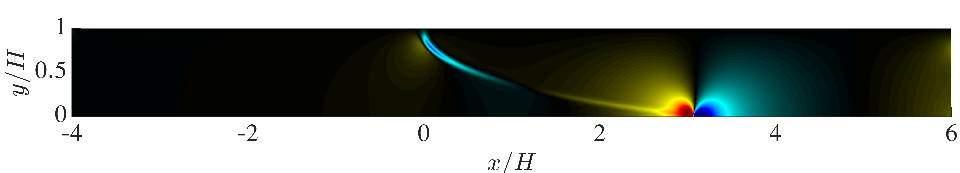}
        \put(2,17){(\textit{b})}
        \end{overpic}
        \vspace{0pt}
        \begin{overpic}[width=1\textwidth]{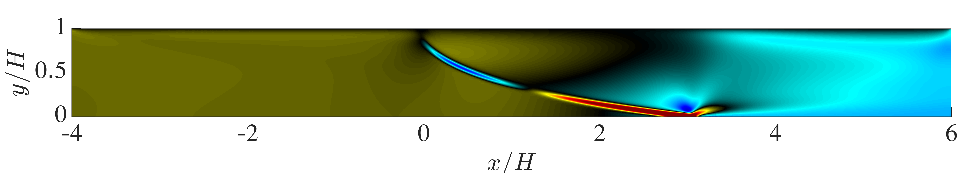}
        \put(2,17){(\textit{c})}
        \end{overpic}        
        \phantomsubcaption        
    \end{subfigure}%
    \hspace{5pt}                       
    \begin{subfigure}[b]{0.08\textwidth}
\includegraphics[width=\textwidth,height=0.32\textheight]{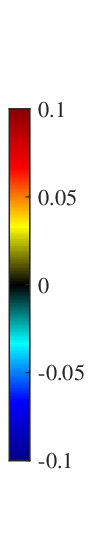}
    \phantomsubcaption
    \end{subfigure}  
  \caption{Streamwise velocity field and Helmholtz decomposition of pure ITA mode in case L10M1. Real part of streamwise velocity field $\hat{u}_{x}$ (\textit{a}), potential part $\hat{u}_{d,x}$ (\textit{b}) and solenoidal part $\hat{u}_{s,x}$ (\textit{c}). All results are normalized by the maximum of $\left| \hat{u}_{x} \right| $.}
\label{fig:pureITA_L10_Helm}
\end{figure}

We delve to characterize the flow features of the pure ITA mode. The mode shapes of heat release rate, characteristic wave and decomposed velocity are presented in figures~\ref{fig:pureITA_L10_Q}, \ref{fig:pureITA_L10_ACwave} and \ref{fig:pureITA_L10_Helm}, respectively. As shown, the induced heat release rate are concentrated in the flame region where the chemical reaction takes place. The flame exhibits strong wrinkles around the tip where the fluctuation amplitude peaks. Those features are similar to the results in the previous study~\citep{varillon2023global}. The heat release fluctuation serves as the major source for acoustic waves, as revealed in figure~\ref{fig:pureITA_L10_ACwave}. In particular, the backward-travelling acoustic wave $\hat{g}$ is able to modulate velocity fluctuation upstream from the flame (see the figure~\ref{fig:pureITA_L10_Helm}(a)), which in return leads to perturbation of heat release rate. This feedback loop observed from the mode features resembles the key characteristics of ITA loop analyzed based on the simplified modes~\citep{bomberg2015thermal,silva2023intrinsic}. 

In order to gain more insights into the modality of this eigenmode and the associated physical mechanism, the Helmholtz decomposition is applied to the velocity field of the pure ITA mode. The results of the Helmholtz decomposition are provided in figure~\ref{fig:pureITA_L10_Helm}(b,c), corresponding to a potential component, $\hat{u}_{d,x}$ and a solenoidal component, $\hat{u}_{s,x}$, respectively. The potential component, $\hat{u}_{d,x}$, is mainly located near the flame front. It does not undergo notable changes in magnitude across the flame and meanwhile exhibits a dipole structure near the flame tip, where the flow field has the largest dilatation effect. The solenoidal component, $\hat{u}_{s,x}$, accounts for the dominant part of velocity fluctuations in terms of magnitude, especially upstream and downstream away from the flame. In the light of flame transfer function, it is noteworthy that the hydrodynamic component, $\hat{u}_{s,x}$, serves as the major contributor to the heat release rate fluctuations of flame, thereby playing a more prominent role in triggering intrinsic thermoacoustic instability. Another interesting feature of the pure ITA mode is that its phase changes across the flame, especially the solenoidal component $\hat{u}_{s,x}$, which indicates the convection attribute of the velocity perturbations acting on the flame front. Besides, larger vorticity emerges around the flame tip, as observed in figure~\ref{fig:pureITA_L10_Helm}(c). 

\subsection{Intrinsic thermoacoustic mode with reflective boundary conditions}

 \begin{figure}
  \centerline{\includegraphics{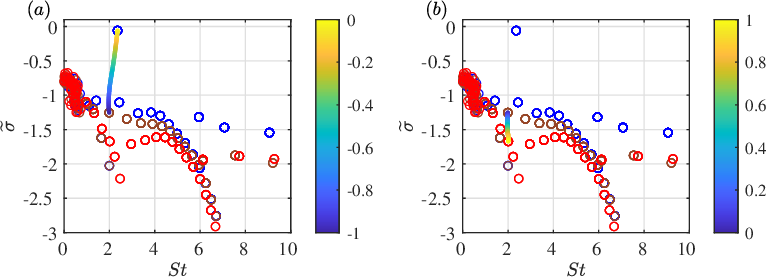}}
  \caption{Trajectories of ITA mode at in case L10M1. (\textit{a}) $R_{out}$ from 0 to -1 at $R_{in}=0$ and (\textit{b}) $R_{in}$ from 0 to 1 at $R_{out}=-1$. Eigenspectra with different boundary conditions are marked by: $R_{in}=R_{out}=0$ (blue $\circ$ ), $R_{in}=0,R_{out}=-1$ (brown $\circ$ ) and $R_{in}=1,R_{out}=-1$ (red $\circ$ ). The color of the trajectories corresponds to the value of the reflection coefficients during the parameter variation.}
\label{fig:trajectory_ITA_l10}
\end{figure}

Now we investigate the influence of reflective boundary conditions on the ITA mode. For this purpose, the parameter variation of reflection coefficients, $R_{in}$ and $R_{out}$, is carried out. This is executed by changing $R_{in}$ from 0 to 1 and $R_{out}$ from 0 to -1 and we track specific eigenmode born of pure ITA mode. The trajectories of ITA mode are shown in figure~\ref{fig:trajectory_ITA_l10}. As shown, the ITA mode becomes much more stable as its growth rate decay during switching the outlet boundary condition from non-reflective to open-end $R_{out}=-1$ with $R_{in}=0$. This has been observed in transient simulations~\citep{silva2023intrinsic} and linear stability analysis~\citep{varillon2023global}. With $R_{in}$ varies from 0 to 1, the growth rate of the ITA mode continues to decay while frequency remains nearly affected. There exist some eigenmodes which are not present in the eigenspectrum of traditional acoustic network models. They are expected to be associated with the hydrodynamic effects, such as the eigenmodes clustering in the very low frequency range. These modes exist irrespective of the acoustic boundary condition. 

\begin{figure}
    \centering
    \begin{subfigure}[h]{0.82\textwidth}
        \begin{overpic}[width=1\textwidth]{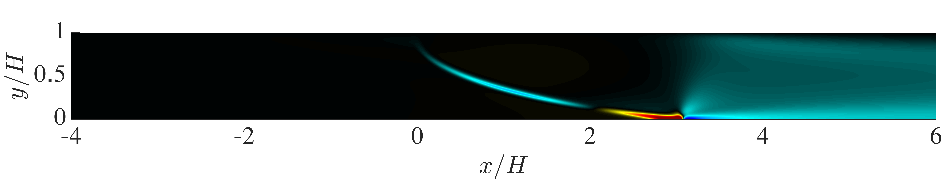}
        \put(2,17){(\textit{a})}
        \end{overpic}    
        \vspace{0pt}
        \begin{overpic}[width=1\textwidth]{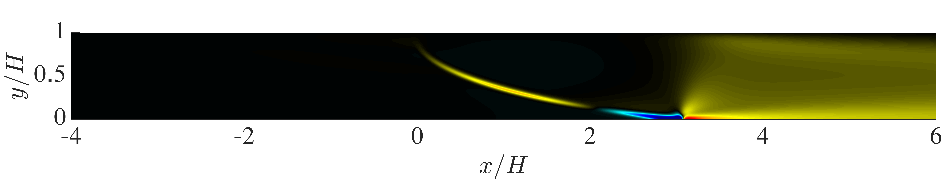}
        \put(2,17){(\textit{b})}
        \end{overpic}
        \phantomsubcaption
    \end{subfigure}%
    \hspace{5pt}                       
    \begin{subfigure}[h]{0.09\textwidth}
\includegraphics[width=\textwidth,height=0.225\textheight]{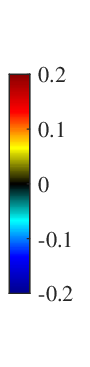}
    \phantomsubcaption
    \end{subfigure}  
  \caption{Acoustic characteristic wave of ITA mode with fully reflective boundary conditions in case L10M1. Real part of forward acoustic characteristic wave $\hat{f}$ (\textit{a}). Real part of backward acoustic characteristic wave $\hat{g}$ (\textit{b}). All results are normalized by its corresponding maximum of $\left| \hat{f} \right| $ or $\left| \hat{g} \right| $.}
\label{fig:ITA_reflective_L10_ACwave}
\end{figure}

\begin{figure}
    \centering
    \begin{subfigure}[h]{0.82\textwidth}
        \begin{overpic}[width=1\textwidth]{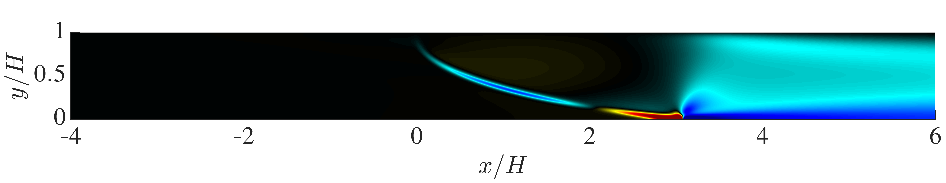}
        \put(2,17){(\textit{a})}
        \end{overpic}    
        \vspace{0pt}
        \begin{overpic}[width=1\textwidth]{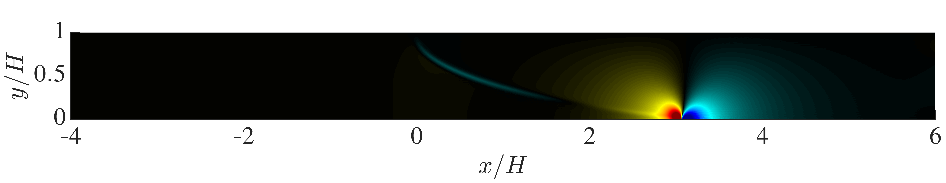}
        \put(2,17){(\textit{b})}
        \end{overpic}
        \vspace{0pt}
        \begin{overpic}[width=1\textwidth]{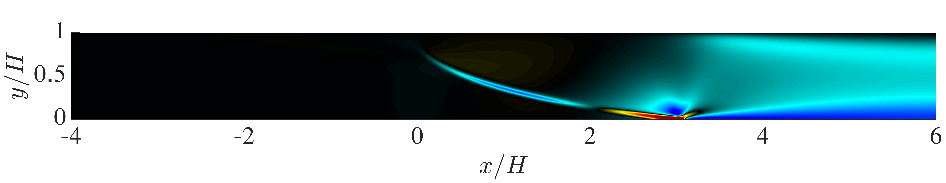}
        \put(2,17){(\textit{c})}
        \end{overpic}        
        \phantomsubcaption        
    \end{subfigure}%
    \hspace{5pt}                       
    \begin{subfigure}[h]{0.08\textwidth}
\includegraphics[width=\textwidth,height=0.32\textheight]{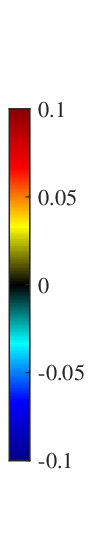}
    \phantomsubcaption
    \end{subfigure}  
  \caption{Streamwise velocity field and Helmholtz decomposition of ITA mode with fully reflective boundary conditions in case L10M1. Real part of streamwise velocity field $\hat{u}_{x}$ (\textit{a}), potential part $\hat{u}_{d,x}$ (\textit{b}) and solenoidal part $\hat{u}_{s,x}$ (\textit{c}). All results are normalized by the maximum of $\left| \hat{u}_{x} \right| $.}
\label{fig:ITA_reflective_L10_Helm}
\end{figure}

We further investigate the mode shapes under acoustically reflective boundary conditions, focusing on the influence of these boundary conditions on the ITA mode and the underlying physical mechanisms. Figures~\ref{fig:ITA_reflective_L10_ACwave} and \ref{fig:ITA_reflective_L10_Helm} present the mode shapes of the acoustic characteristic waves and the streamwise velocity fluctuations, along with their Helmholtz decomposition. The current configuration corresponds to a closed-open combustor. It is observed that this configuration produces a much weaker backward acoustic wave, thereby diminishing the modulation effect of acoustic waves on the upstream velocity field. This behavior contrasts sharply with the scenario observed in the anechoic configuration for the pure ITA mode. In the fully reflective configuration ($R_{in}=1$ and $R_{out}=-1$), the ITA mode exists within a significantly damped ITA loop, resulting in a much smaller growth rate. The Helmholtz decomposition of the fluctuating velocity field, shown in the figure~\ref{fig:ITA_reflective_L10_Helm}, reveals that, despite the notable changes in magnitude, particularly in the solenoidal component, $\hat{u}_{s,x}$, the decomposed velocity fields exhibit very similar structures and localized features to those of the pure ITA mode, as shown in figure~\ref{fig:pureITA_L10_Helm}. This indicates that the intrinsic physical processes governing the ITA mode remain unchanged despite variations in acoustic boundary conditions. Consequently, the features revealed through Helmholtz decomposition may serve to identify thermoacoustic modes that originate from ITA modes. 

\section{Identification of exceptional point}
\label{identification exceptional points}
In this section, we perform parameter variation to examine the behaviors of thermoacoustic mode trajectories and mode interplay between ITA and AC modes in the proximity of an exceptional point. The results are obtained by considering different choices of parameters and different paths of parameter variations. 

\subsection{Methodology--parameter variation}
Exceptional points frequently emerge at spectral locus where two  different branches of eigenmodes collide. To investigate such interactions, it is essential to focus on configurations where the ITA mode closely aligns in frequency with a specific AC mode, which is a necessary condition for mode interplay. Traditionally, parameter variation has been widely used to explore different configurations and cases, allowing for the tracking of mode trajectories and their interactions. This technique has proven effective in identifying the origins of typical eigenmodes and exceptional points. In previous studies, parameter variations have been performed on combustor chamber length, acoustic reflection coefficients, an artificial coupling factor~\citep{ghani2021exceptional}, and parameters in simplified flame models, for example, the $n-\tau$ model~\citep{sogaro2019thermoacoustic}. However, the artificial coupling factor is a physically virtual parameter and parameters in the $n-\tau$ model may not be physically realizable. Here we use length of combustor and inflow Mach number to adjust the frequency of AC mode. The sound speed is lowered with inflow Mach number increased. The variation of inflow Mach number will not only be incorporated into the baseflow solutions, but also directly reflected on the form of acoustic eigenvalue problems. Considering a typical closed-open combustor setup with $R_{in}=1$ and $R_{out}=-1$, the first AC mode corresponds to a quarter-wavelength mode and the non-dimensional frequency could be estimated without considering the temperature increment across the flame and the fluctuations of heat release rate due to the unsteady combustion process,
\begin{equation}
St_{AC, 1st} = \frac{ \omega H}{u_{0}} \approx \frac{ (\frac{4 c_{0}}{L}) \times H} {u_{0}} = \frac{4}{ (\frac{L}{H}) \times Ma} \;,
\end{equation}
where $u_0$ and $c_{0}$ corresponds to the flow speed and the speed of the sound at the inflow. A more accurate estimation on AC modes can be enabled by solving the Helmholtz equations~\citep{nicoud2007acoustic}. By omitting the baseflow velocity and incorporating the Mach number into the non-dimensional form, the Helmholtz equations can be expressed as:
\begin{equation}
	\gamma (\overline{p}+\frac{1}{\gamma Ma^{2}}) \bnabla \cdot (\frac{1}{\overline{\rho}} \bnabla \hat{p}) + St^{2}\hat{p}=0 \;.
\label{eqn:hel_for_AC}
\end{equation}
It should be noted that herein the inflow Mach number, $Ma$, serves a similar function to the acoustic duct length in terms of modifying acoustic frequencies. However, the Mach number is a parameter directly related to the operating conditions of the combustor and is more practical to adjust compared to the physical length of the combustor. 
The influence of inflow Mach number on theromoacoustic instability has been studied with acoustic network models~\citep{nan2022theoretical}. However, to the best of our knowledge, the impact of different inflow Mach numbers on the existence of exceptional point in the framework of compressible reactive flow equations has not yet been investigated and discussed. 

\begin{table}
  \begin{center}
\def~{\hphantom{0}}
  \begin{tabular}{ccccc}
     Case & Parameter  & Marker & First AC mode & Second AC mode \\[3pt]
    L10M1 &   $L/H=10$, Ma = 0.01 & $\circ$  & 29.31 & 71.15  \\
    L50M1 &  $L/H=50$, Ma = 0.01  & $\triangle$ & 6.12  & 14.31 \\
    L50M2 &   $L/H=50$, Ma = 0.02 & $\square$  & 3.06  & 7.15 \\
    L50M3 &  $L/H=50$, Ma = 0.03  & $\triangledown$ & 2.04  & 4.77 \\
  \end{tabular}
  \caption{Frequencies of pure AC modes estimated by the Helmholtz equations. All results are non-dimensionalized as $St = \omega H/u_{0}$.}
  \label{tab:solution of helmholtz system}
  \end{center}
\end{table}

In the present study, we primarily consider the inflow Mach and acoustic reflection coefficients, $R_{in}$ and $R_{out}$, to identify the exceptional point and analyze the interplay between ITA and AC mode around it. A heat release factor $n$ is also supplemented to help identifying the AC mode origin. This is enabled by multiplying the heat release factor to the chemical source term $\hat{\dot{\omega}}_T$. As such, the case of $n=0$ refers to a passive flame scenario. Several distinct case setups that are the focus of the analysis are listed in table~\ref{tab:solution of helmholtz system}. Case L10M1 with $L/H=10$ and $Ma=0.01$ has been utilized to identify the ITA mode in~\cref{Intrinsic thermoacoustic mode} and will serve as a reference in pinpointing the ITA mode in the following analysis. We resort to the configuration with a longer duct and a higher Mach number to analyze the mode interplay, as it brings the frequency of first AC mode closer to the frequency of identified ITA mode. Here we solving equation~(\ref{eqn:hel_for_AC}) to obtain the AC modes with passive flame limit for different cases. The first and second AC mode frequencies are given in table~\ref{tab:solution of helmholtz system}. It is evident that extending the duct length and increasing Mach number both effectively reduces the AC mode frequencies. Nevertheless, this fact also qualifies the Mach number to be a candidate for parameter variation. 

\begin{figure}
  \centerline{\includegraphics{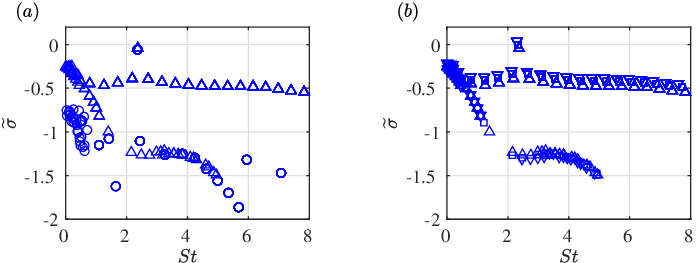}}
  \caption{Eigenspectra for different duct lengths and Mach numbers with $R_{in}=R_{out}=0$. (\textit{a}) Comparison for different duct lengths at $Ma=0.01$ and (\textit{b}) comparison for different Mach numbers at $L/H=50$. Eigenspectra are marked by: case L10M1 (blue $\circ$ ), L50M1 (blue $\triangle$ ), L50M2 (blue $\square$ ) and L50M3 (blue $\triangledown$ ).}
\label{fig:ev_compare_pureITA}
\end{figure}

A critical consideration in selecting parameters is ensuring that the pure ITA mode remains largely unaffected during parameter variations. This approach allows for the tuning of the acoustic mode frequency to induce mode interplay. To this end, we compute and compare the eigenspectra for the cases considered under an anechoic configuration, with the results presented in figure~\ref{fig:ev_compare_pureITA}. As expected, the pure ITA mode remains unchanged across different $L/H$ ratios, given that the physical mechanism governing the ITA mode is localized near the flame. In contrast, certain low-frequency eigenmodes, such as vortical modes, exhibit variations that can be attributed to their sensitivity to the length of the computational domain. Regarding the influence of the Mach number, the pure ITA mode is found to be only slightly destabilized as $Ma$ increases. At low Mach numbers, the flame dynamics are predominantly governed by near-field hydrodynamics, with minimal influence from Mach number effects, a well-established concept in the numerical combustion community, as demonstrated in previous studies~\citep{emmett2014high}. This concept is also supported by prior research comparing compressible and incompressible simulations to identify flame dynamics~\citep{eder2023incompressible}. Consequently, within the low Mach number range investigated here, the pure ITA mode can be considered nearly invariant with respect to Mach number. This property qualifies the inflow Mach number, $Ma$, as a suitable parameter for variations.

\subsection{Mode trajectories and mode origins}

\begin{figure}
  \centerline{\includegraphics{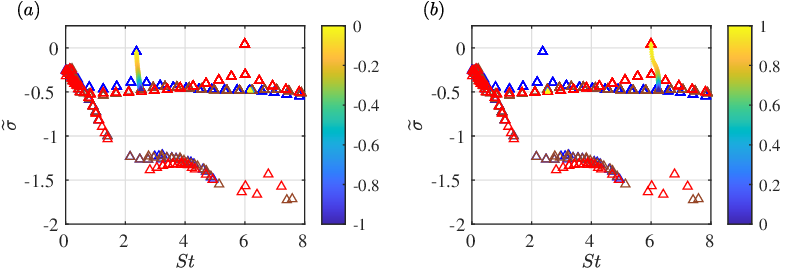}}
  \caption{Trajectories of ITA mode and AC mode in case L50M1. (\textit{a}) $R_{out}$ from 0 to -1 at $R_{in}=0$ and (\textit{b}) $R_{in}$ from 0 to 1 at $R_{out}=-1$. Eigenspectra with different boundary conditions are marked by: $R_{in}=R_{out}=0$ (blue $\triangle$ ), $R_{in}=0,R_{out}=-1$ (brown $\triangle$ ) and $R_{in}=1,R_{out}=-1$ (red $\triangle$ ). The color of the trajectories corresponds to the value of the reflection coefficients during the parameter variation.  }
\label{fig:trajectory_ITA_l50_Ma0.01}
\end{figure}

\begin{figure}
  \centerline{\includegraphics{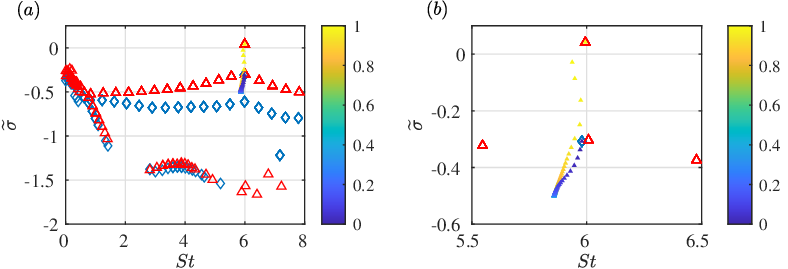}}
  \caption{Trajectories of AC mode with fully non-reflective boundaries in case L50M1. (\textit{a}) $n$ from 0 to 1 at $R_{in}=1,R_{out}=-1$ and (\textit{b}) a magnified view of (\textit{a}) focused on the AC mode. Eigenspectra with $n$ are marked by: $n=1$ (red $\triangle$ ) and $n=0$ (light blue $\Diamond$ ). The color of the trajectories corresponds to the value of $n$ during the parameter variation. }
\label{fig:trajectory_n_ac_l50_Ma0.01}
\end{figure}

For each case at a given inflow Mach number, parameter variation is conducted with respect to the boundary reflection coefficients following a specific path: $R_{out}$ is varied from 0 to -1 with $R_{in}=0$, followed by varying $R_{in}$ from 0 to 1 with $R_{out}=-1$. This approach transitions the duct configuration from anechoic to fully reflective conditions. The results for case L50M1 are depicted in figure~\ref{fig:trajectory_ITA_l50_Ma0.01}. The growth rate of the ITA mode decreases during the parameter variation, while an unstable eigenmode emerges at a higher frequency when the duct becomes fully reflective. This unstable eigenmode is anticipated to originate from the AC mode, a hypothesis we verify by examining the heat-release parameter $n$. Figure~\ref{fig:trajectory_n_ac_l50_Ma0.01} presents the eigenspectra at various values of $n$, along with the mode trajectory as $n$ varies. By tracing this trajectory, the least stable eigenmode is identified at $n=0$, with a frequency very close to that of the quarter-wavelength mode at $St=6.12$. This finding indicates that the dominant unstable eigenmode at $n=1$ indeed originates from an AC mode.
 
\begin{figure}
  \centerline{\includegraphics{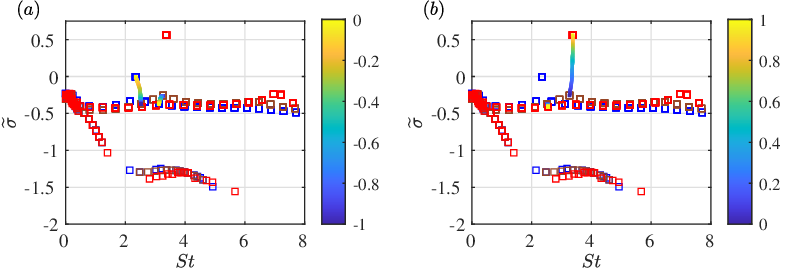}}
  \caption{Trajectories of ITA mode and AC mode in case L50M2. (\textit{a}) $R_{out}$ from 0 to -1 at $R_{in}=0$ and (\textit{b}) $R_{in}$ from 0 to 1 at $R_{out}=-1$. The color of markers in the eigenspectra and color of the trajectories are the same with figure \ref{fig:trajectory_ITA_l50_Ma0.01}.}
\label{fig:trajectory_ITA_l50_Ma0.02}
\end{figure}

\begin{figure}
  \centerline{\includegraphics{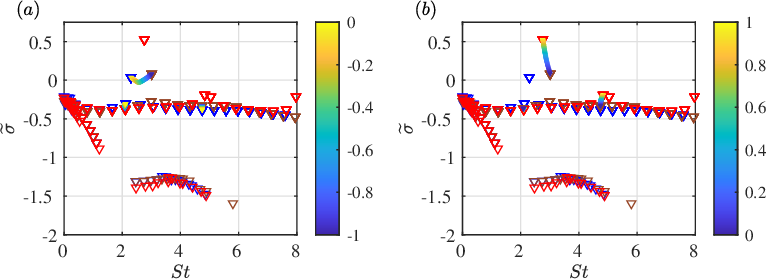}}
  \caption{Trajectories of ITA mode and AC modes in case L50M3. (\textit{a}) $R_{out}$ from 0 to -1 at $R_{in}=0$ and (\textit{b}) $R_{in}$ from 0 to 1 at $R_{out}=-1$. The color of markers in the eigenspectra and color of the trajectories are the same with figure \ref{fig:trajectory_ITA_l50_Ma0.01}.}
\label{fig:trajectory_ITA_l50_Ma0.03}
\end{figure}

We proceed to examine cases L50M2 and L50M3 through parameter variation. In these two cases, the frequencies of the ITA and AC modes are in much closer proximity, raising the possibility of mode coalescence in the presence of an exceptional point. The results for cases L50M2 and L50M3 are presented in figures~\ref{fig:trajectory_ITA_l50_Ma0.02} and \ref{fig:trajectory_ITA_l50_Ma0.03}, respectively. In case L50M2, the eigenspectrum and its variation closely resemble those in case L50M1, where the unstable thermoacoustic mode originates from the AC mode while the ITA mode is stabilized. However, in case L50M3, a branch switching phenomenon occurs. The unstable thermoacoustic mode originates from the ITA mode, with the first and second AC modes becoming damped. The trajectory of the ITA mode veers and changes direction before stabilizing, underscoring the presence of an exceptional point in this thermoacoustic system.

\begin{figure}
  \centerline{\includegraphics{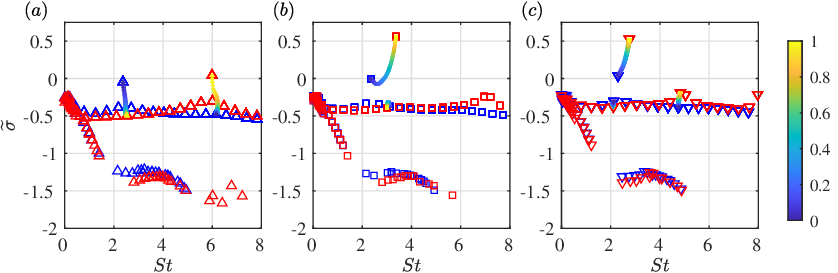}}
  \caption{Trajectories of ITA mode and AC modes during parameter variation for $R_{k}$ from 0 to 1 in case L50M1, L50M2 and L50M3. The color of markers in the eigenspectra: non-reflective boundary conditions of $R_{k}=0$ (dark blue) and fully-reflective boundary conditions of $R_{k}=1$ (red). The color of the trajectories corresponds to the value of $R_{k}$ during the parameter variation.}
\label{fig:trajectory_Rk}
\end{figure}

To study the impact of the exceptional point in more details, we limit the number of parameters to two: the inflow Mach numebr $Ma$ and a auxiliary parameter $R_{k}$ to indicate the change of boundary conditions from non-reflective ($R_{in}=R_{out}=0$) to fully-reflective ($R_{in}=1, R_{out}=-1$). $R_{k}$ is defined as $R_{in}=-R_{out}=R_{k}$, varying from 0 to 1. As a result of parameter variation, the eigenspectra and mode trajectories for the three target cases are presented in figure~\ref{fig:trajectory_Rk}. Comparing figure~\ref{fig:trajectory_Rk}(a) to figure~\ref{fig:trajectory_ITA_l50_Ma0.01} and figure~\ref{fig:trajectory_Rk}(c) to figure~\ref{fig:trajectory_ITA_l50_Ma0.03}, we find that the mode trajectories and the identified mode origins are consistent for cases L50M1 and L50M3, regardless of the parameter path. The event of trajectory veer is clearly recognized. However, it is remarkable to find that in case L50M2 the results of identified thermoacoustic mode origin are completely different depending on the path of parameter variation, even though the parameter intervals are exactly the same. When the parameters ($R_{in},R_{out}$) vary sequentially from (0,0) to (0,-1) and then to (1,-1), as shown in figure~\ref{fig:trajectory_ITA_l50_Ma0.02}, the unstable thermoacoustic mode is found to be born of a AC mode. On the other hand, when the parameters, $R_{in}$ and $R_{out}$, change simultaneously along the line ($R_{in}=-R_{out}$), the unstable thermoacoustic mode originates from the ITA mode, as revealed in figure~\ref{fig:trajectory_Rk}(b). This inconsistency is attributed to the high parametric sensitivity near the exceptional point~\citep{mensah2018exceptional,schaefer2021impact}. When the parameter variation is conducted numerically, even with very small parameter increment for each step, the solutions can be strongly influenced by the existence of exceptional point. It implies that the caution should be paid to the identification of thermoacoustic mode near the exceptional point by parameter variation.

\subsection{Locus of exceptional point and mode interplay} 
\begin{figure}
  \centerline{\includegraphics[width=0.7\textwidth]{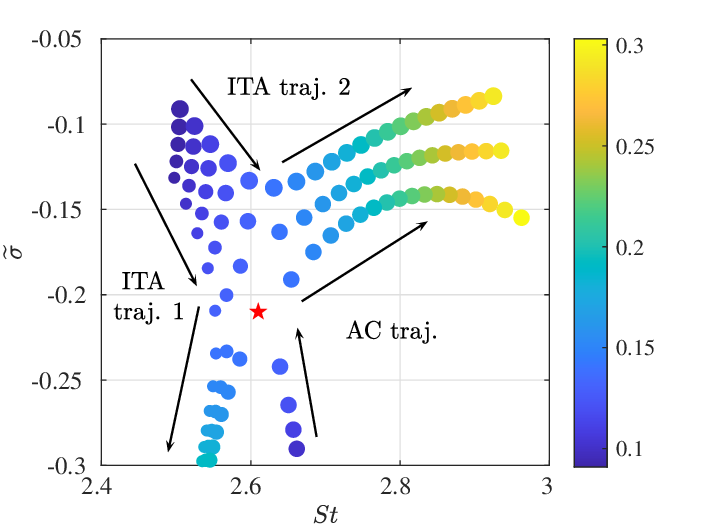}}
  \caption{Detection of the exceptional point with $Ma$ and $R_{k}$ at $L/H=50$. The size and color of markers are used to denote the values of $Ma$ and $R_k$, respectively. Here the shown parameter ranges are $Ma \in [0.158,~0.184]$ and $R_{k} \in [0.1,~0.3]$. The arrows are used to indicate mode trajectories during parameter variation. The exceptional point is marked with a red pentagram. }
\label{fig:trajectory_Ma_Rk}
\end{figure}

The veering of mode trajectories, as shown in figure~\ref{fig:trajectory_Rk}, suggests the presence of exceptional points near the parameter set of case L50M2. To precisely locate the exceptional point on the eigenspectrum, we incrementally varied the parameter values in small steps. For each set of parameters, the eigenmode trajectories were computed and used to determine the locus of the exceptional point. Through this process, we numerically identified the exceptional point corresponding to a critical parameter set around $Ma = 0.0174$ and $R_{k} = 0.14$. The associated defective eigenvalue was found to be $\widetilde{\sigma} = -0.21$ with a Strouhal number $St = 2.61$. A magnified view of the mode trajectories near the exceptional point is presented in figure~\ref{fig:trajectory_Ma_Rk}. As depicted, the trajectories of the ITA mode and AC mode coalesce at the exceptional point. For Mach numbers below the critical value, the ITA mode trajectory extends downward in a vertical direction. When the Mach number exceeds its critical value, the ITA mode trajectory veers towards the right in a horizontal direction. Meanwhile, the AC mode trajectory approaches the exceptional point from below and then veers to the right. Therefore, an abrupt brach switch is observed here. The unstable thermoacoustic modes originally born of AC modes become to be arising from ITA modes, and vice versa. The mode interplay behavior analyzed here is consistent with findings from previous studies in the framework of acoustic network models with other parameters~\citep{sogaro2019thermoacoustic,ghani2021exceptional}, demonstrating the effectiveness of the biglobal linear stability approach in investigating mode interplay near exceptional points.

\begin{figure}
  \centerline{\includegraphics[width=0.7\textwidth]{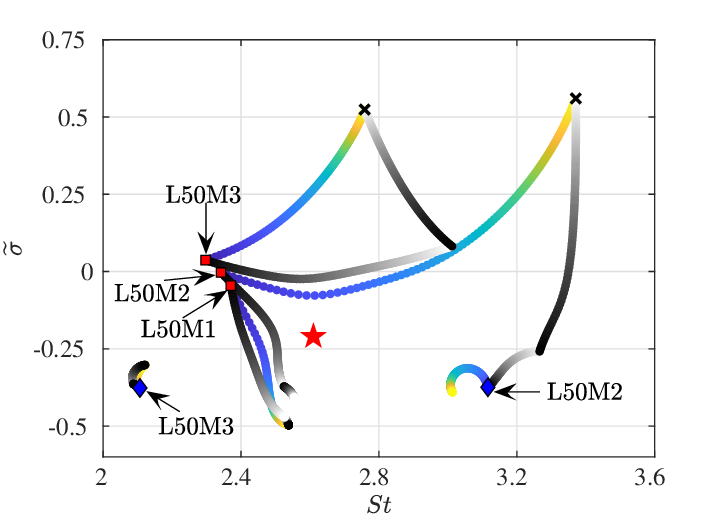}}
  \caption{The trajectories of eigenmodes during parameter variation with different paths. The pure ITA modes and AC modes are marked by red $\square$ and blue $\Diamond$. The unstable thermoacoustic modes are marked by black $\times$. The color of trajectories of path 1 corresponds to the value of $R_{k}$ with Parula colormap and the color of trajectories of path 2 corresponds to the value of $R_{in}$ and $R_{out}$ with Gray colormap. The exceptional point is highlighted by a red pentagram. }
\label{fig:all_trajectories with ep}
\end{figure}

It has been recognized that the Mach number of case L50M2 is very close to that of the identified exceptional point. The high parametric sensitivity in this case therefore is likely attributed to inconsistency in identified mode origin. We revisit the eigenmode trajectories of the examined cases following different paths of parameter variations, which are shown in figure~\ref{fig:all_trajectories with ep}. The two paths considered are consistent with the previous analysis. Mathematically, path 1 corresponds to $R_{in}$ and $R_{out}$ varying simultaneously with $R_{in}=-R_{out}$, while path 2 corresponds to $R_{in}$ and $R_{out}$ varying separately from (0,~0) to (0,~-1), then to (1,~-1). For case L50M1, the eigenmode trajectories of the two parametric paths are issued from the ITA origin and running side-by-side downward, and eventually reaches the same stable eigenmode. The trajectories in case L50M3, in contrast, are both running toward the same unstable eigenmode. However, in case L50M3 the dependence of trajactory on parametric path is clearly visualized as the two trajectories issued from the ITA origin diverge initially, although finishing at the same end-point. Case L50M2 is where the abnormal mode behavior is identified. As shown, the two mode trajectories, which are born of the pure ITA mode, diverges with respect to the exceptional point and run towards two different sides. The two mode trajectories no longer share the same end-point. At this point, it is in fact not clear whether the unstable eigenmode in that case emerges from ITA or AC origin, based solely on the parameter variation. Previous studies have recognized the dependence of identified mode origin on the choice of parameter~\citep{silva2023intrinsic}. Here we specifically show that even for the same parameter, the results on identified mode can differ case-by-case, depending on how parameter varies technically. These results further substantiate that in the vicinity of exceptional point, mode origin identification based on parameter variation could be highly unreliable. It becomes numerically inaccurate to distinguish trajectories of different types of thermoacoustic modes with solely conducting parameter variation. Therefore, it is critical to identify thermoacoustic modes by another means that has less parametric dependence. Inspired by the results demonstrated in~\cref{Intrinsic thermoacoustic mode} that the characteristic mode structures of ITA mode extracted with Helmholtz decomposition remain nearly unchanged when the boundary conditions vary from anechoic to fully reflective, we will further investigate the mode shapes of thermoacoustic modes born of different types of eigenmodes and apply this to help us categorize thermoacoustic modes. Besides, the features of eigenmodes in the proximity of exceptional point will be our focus to reveal the influence of exceptional point in the mode shapes of thermoacoustic modes.

\section{Characteristic mode structures}
\label{Mode structure}
As the parameter variation could be rather tedious and suffer from robustness issue in categorizing mode origin, we therefore tend to rely on a more straightforward and physical based approach to enable categorization of thermoacoustic modes. It might be better to categorize thermoacoustic modes by their physical features instead of by parameter variation, and see if there are key characteristic mode structure that may discriminate different thermoacoustic modes. 

\subsection{Mode shapes for different origins}
\begin{figure}
    \centering
    \begin{subfigure}[b]{0.42\textwidth}
        \centering
        \begin{overpic}[width=1\textwidth]{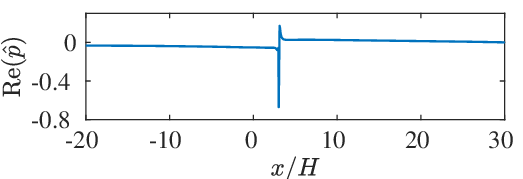}
        \put(2,35){(\textit{a})}
        \end{overpic}  
        \phantomsubcaption 
    \end{subfigure}
    \hspace{0.2pt}
    \begin{subfigure}[b]{0.42\textwidth}
        \centering
        \begin{overpic}[width=1\textwidth]{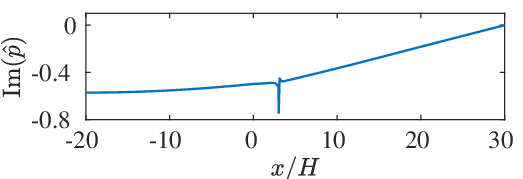}
        \put(2,35){(\textit{b})}
        \end{overpic}  
        \phantomsubcaption 
    \end{subfigure}    
    \vspace{0.1pt} 
    \begin{subfigure}[b]{0.42\textwidth}
        \centering
        \begin{overpic}[width=1\textwidth]{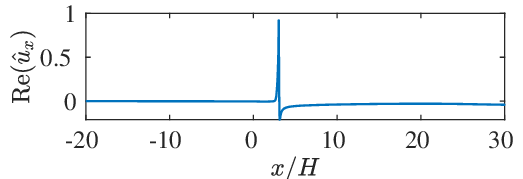}
        \put(2,35){(\textit{c})}
        \end{overpic}  
        \phantomsubcaption 
    \end{subfigure}
    \hspace{0.2pt} 
    \begin{subfigure}[b]{0.42\textwidth}
        \centering
        \begin{overpic}[width=1\textwidth]{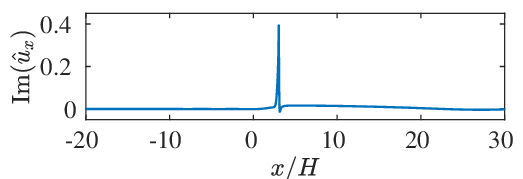}
        \put(2,35){(\textit{d})}
        \end{overpic}  
        \phantomsubcaption 
    \end{subfigure}
\caption{Pressure and streamwise velocity profiles of the thermoacoustic mode originating from ITA mode in case L50M1. Pressure on the central line (\textit{a},\textit{b}). Streamwise velocity on the central line (\textit{c},\textit{d}). Real part (\textit{a},\textit{c}) and imaginary part (\textit{b},\textit{d}). All results are normalized by its corresponding maximum of $\left| \hat{p} \right| $ or $\left| \hat{u}_{x} \right| $ on the central line.}
\label{fig:Global_ITA_l50_Ma0.01}
\end{figure}

\begin{figure}
    \centering
    \begin{subfigure}[b]{0.82\textwidth}
        \begin{overpic}[width=1\textwidth]{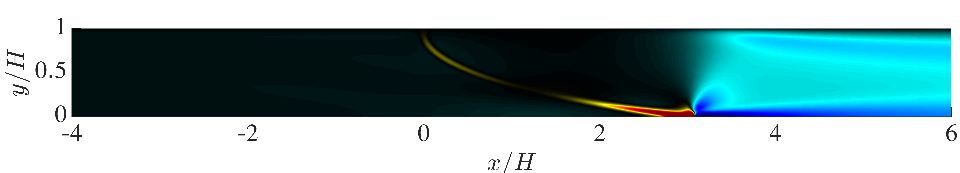}
        \put(2,17){(\textit{a})}
        \end{overpic}    
        \vspace{0pt}
        \begin{overpic}[width=1\textwidth]{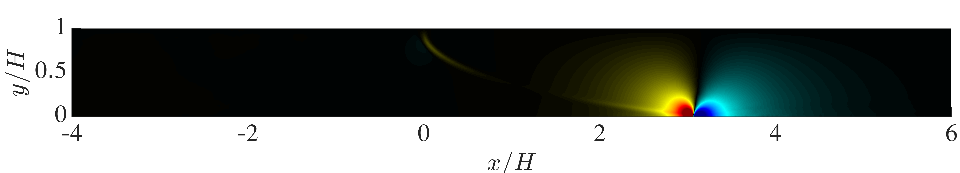}
        \put(2,17){(\textit{b})}
        \end{overpic}
        \vspace{0pt}
        \begin{overpic}[width=1\textwidth]{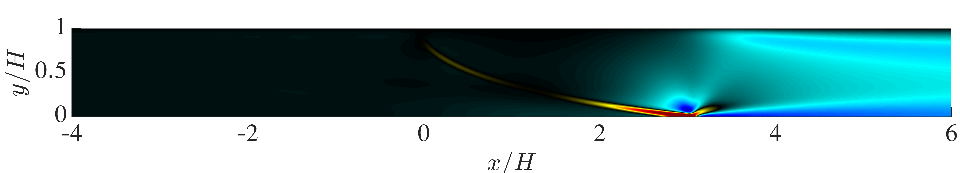}
        \put(2,17){(\textit{c})}
        \end{overpic}        
        \phantomsubcaption        
    \end{subfigure}%
    \hspace{5pt}                       
    \begin{subfigure}[b]{0.1\textwidth}
\includegraphics[width=\textwidth,height=0.32\textheight]{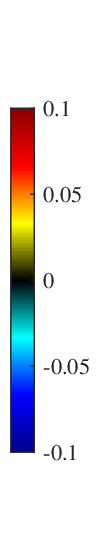}
    \phantomsubcaption
    \end{subfigure}    
\caption{Real part of streamwise velocity field and Helmholtz decomposition of thermoacoustic mode born of ITA mode with fully reflective boundary conditions in case L50M1. Streamwise velocity field $\hat{u}_{x}$ (\textit{a}), potential component $\hat{u}_{d,x}$ (\textit{b}) and solenoidal component $\hat{u}_{s,x}$ (\textit{c}). All results are normalized by the maximum of $\left| \hat{u}_{x} \right| $.}
\label{fig:ITA_reflective_l50_Ma0.01_Helm}
\end{figure}

We begin by examining the characteristics of the ITA mode under fully reflective boundary conditions in case L50M1. The global profiles of pressure and streamwise velocity along the centerline of the acoustic duct are presented in figure~\ref{fig:Global_ITA_l50_Ma0.01}. Although the acoustic fields near the flame are non-planar, the centerline profiles are chosen to simplify the illustration of the different thermoacoustic modes. The thermoacoustic mode originating from the ITA mode is primarily characterized by its behavior in the flame region, where a sharp velocity discontinuity is observed across the flame. The streamwise velocity fields and their Helmholtz decomposed counterparts near the flame are shown in figure~\ref{fig:ITA_reflective_l50_Ma0.01_Helm}. These results closely resemble those observed in case L10M1 (see figure \ref{fig:ITA_reflective_L10_Helm}), which can be attributed to the fact that the thermoacoustic mode arising from the ITA mode is dominated by the hydrodynamic flows in the vicinity of the flame. In this case, only the length of the acoustic duct is increased, which does not significantly affect the hydrodynamic flows near the flame, despite the longer development of the boundary layer downstream. Consequently, the fundamental physical nature of the ITA mode remains largely unchanged and can be effectively applied to identify different thermoacoustic modes.

\begin{figure}
    \centering
    \begin{subfigure}[b]{0.42\textwidth}
        \centering
        \begin{overpic}[width=1\textwidth]{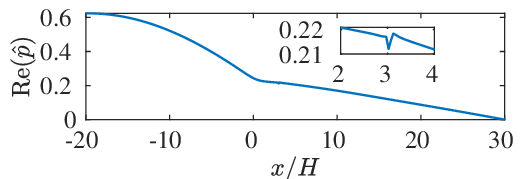}
        \put(2,35){(\textit{a})}
        \end{overpic}  
        \phantomsubcaption 
    \end{subfigure}
    \hspace{0.2pt}
    \begin{subfigure}[b]{0.42\textwidth}
        \centering
        \begin{overpic}[width=1\textwidth]{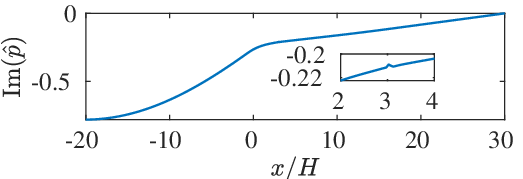}
        \put(2,35){(\textit{b})}
        \end{overpic}  
        \phantomsubcaption 
    \end{subfigure}    
    \vspace{0.1pt} 
    \begin{subfigure}[b]{0.42\textwidth}
        \centering
        \begin{overpic}[width=1\textwidth]{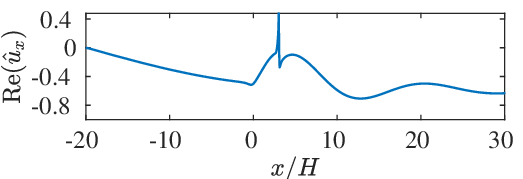}
        \put(2,35){(\textit{c})}
        \end{overpic}  
        \phantomsubcaption 
    \end{subfigure}
    \hspace{0.2pt} 
    \begin{subfigure}[b]{0.42\textwidth}
        \centering
        \begin{overpic}[width=1\textwidth]{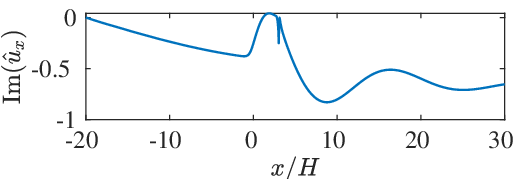}
        \put(2,35){(\textit{d})}
        \end{overpic}  
        \phantomsubcaption 
    \end{subfigure}
\caption{Pressure and streamwise velocity profiles of the thermoacoustic mode originating from AC mode in case L50M1. Pressure on the central line (\textit{a},\textit{b}). Streamwise velocity on the central line (\textit{c},\textit{d}). Real part (\textit{a},\textit{c}) and imaginary part (\textit{b},\textit{d}). All results are normalized by its corresponding maximum of $\left| \hat{p} \right| $ or $\left| \hat{u}_{x} \right| $ on the central line.}
\label{fig:Global_ac_l50_Ma0.01}
\end{figure}

\begin{figure}
    \centering
    \begin{subfigure}[b]{0.82\textwidth}
        \begin{overpic}[width=1\textwidth]{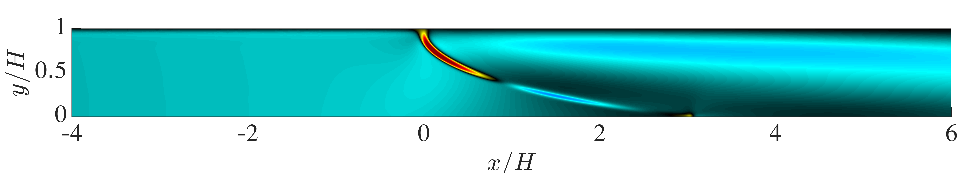}
        \put(2,17){(\textit{a})}
        \end{overpic}    
        \vspace{0pt}
        \begin{overpic}[width=1\textwidth]{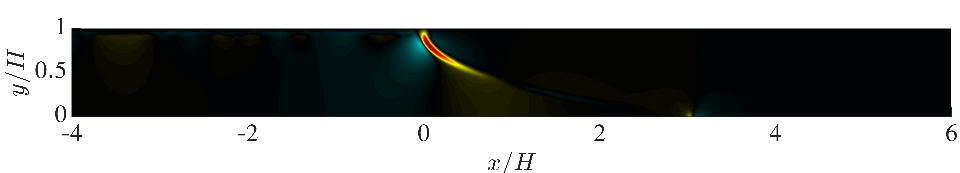}
        \put(2,17){(\textit{b})}
        \end{overpic}
        \vspace{0pt}
        \begin{overpic}[width=1\textwidth]{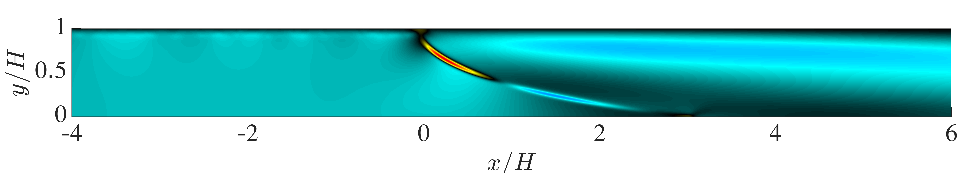}
        \put(2,17){(\textit{c})}
        \end{overpic}        
        \phantomsubcaption        
    \end{subfigure}%
    \hspace{5pt}                       
    \begin{subfigure}[b]{0.1\textwidth}
\includegraphics[width=\textwidth,height=0.32\textheight]{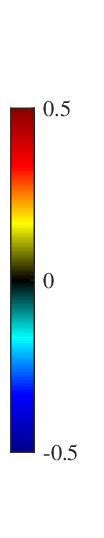}
    \phantomsubcaption
    \end{subfigure}  
\caption{Real part of streamwise velocity field and Helmholtz decomposition of thermoacoustic mode born of AC mode with fully reflective boundary conditions in case L50M1. Streamwise velocity field $\hat{u}_{x}$ (\textit{a}), potential component $\hat{u}_{d,x}$ (\textit{b}) and solenoidal component $\hat{u}_{s,x}$ (\textit{c}). All results are normalized by the maximum of $\left| \hat{u}_{x} \right| $.}
\label{fig:ac_reflective_l50_Ma0.01_Helm}
\end{figure}

The unstable thermoacoustic mode under fully reflective boundary conditions in case L50M1 is identifed to originate from the AC mode. The pressure and streamwise velocity profiles of the eigenmode along the centerline of the acoustic duct are depicted in figure~\ref{fig:Global_ac_l50_Ma0.01}. It is noteworthy that the velocity field is distributed along the entire length of the acoustic duct, rather than being confined to the vicinity of the flame, as is characteristic of modulation by the natural acoustic mode. The pressure profile along the centerline exhibits a weak bump at the flame front, contrasting with the sharp spike observed in figure~\ref{fig:Global_ITA_l50_Ma0.01}(a-b). This suggests that, compared to the ITA mode, the pressure disturbance associated with the AC mode exhibits a pronounced ability to penetrate the flame. The detailed streamwise velocity fields near the flame, along with the results of Helmholtz decomposition, are shown in figure~\ref{fig:ac_reflective_l50_Ma0.01_Helm}. These results differ markedly from those of the thermoacoustic mode originating from the ITA mode, as shown in figure~\ref{fig:ITA_reflective_l50_Ma0.01_Helm}. Unlike the thermoacoustic mode born from the ITA mode, where the primary components of the feedback loop for instability are localized around the flame region, the flame in this case acts more as a localized disturbance within the acoustically modulated global flow fields. The non-planar structure of the flame induces non-planar flow disturbances. The structure of the potential component is primarily concentrated at the flame base, while the solenoidal component is characterized by the development of a shear layer and its interaction with non-planar acoustic waves.

\subsection{Mode shapes near exceptional point}
According to the foregoing analysis, it is evident that the thermoacoustic modes arising from ITA and AC modes exhibit distinct physical characteristics, as manifested in the flow fields and their Helmholtz decomposition. We proceed to investigate the unstable thermoacoustic mode in case L50M2, which lies in close proximity to the identified exceptional point. It is important to note that prior variations in parameters have yielded inconsistent results regarding the origin of this mode. Specifically, parameter variation has not conclusively determined whether the unstable thermoacoustic mode originates from ITA or AC modes. It is therefore of critical importance to ascertain whether this mode exhibits distinct features characteristic of either ITA or AC modes. In addition, the unstable thermoacoustic mode of case L50M3, which is determined to be born of ITA mode by parameter variation, is also presented for comparison. 

\begin{figure}
    \centering
    \begin{subfigure}[b]{0.42\textwidth}
        \centering
        \begin{overpic}[width=1\textwidth]{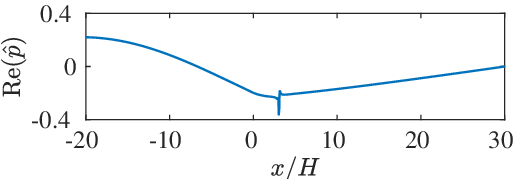}
        \put(2,35){(\textit{a},1)}
        \end{overpic}  
        \phantomsubcaption 
    \end{subfigure}
    \hspace{0.2pt}
    \begin{subfigure}[b]{0.42\textwidth}
        \centering
        \begin{overpic}[width=1\textwidth]{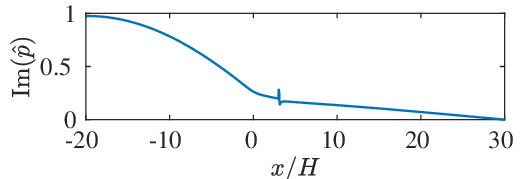}
        \put(2,35){(\textit{a},2)}
        \end{overpic}  
        \phantomsubcaption 
    \end{subfigure}    
    \vspace{0.1pt} 
    \begin{subfigure}[b]{0.42\textwidth}
        \centering
        \begin{overpic}[width=1\textwidth]{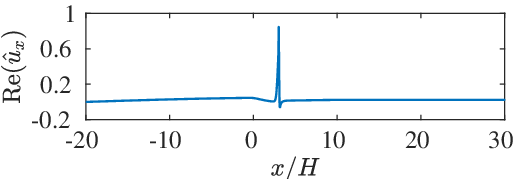}
        \put(2,35){(\textit{b},1)}
        \end{overpic}  
        \phantomsubcaption 
    \end{subfigure}
    \hspace{0.2pt} 
    \begin{subfigure}[b]{0.42\textwidth}
        \centering
        \begin{overpic}[width=1\textwidth]{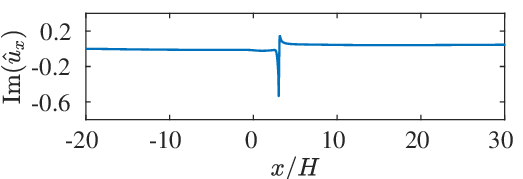}
        \put(2,35){(\textit{b},2)}
        \end{overpic}  
        \phantomsubcaption 
    \end{subfigure}
    \begin{subfigure}[b]{0.42\textwidth}
        \centering
        \begin{overpic}[width=1\textwidth]{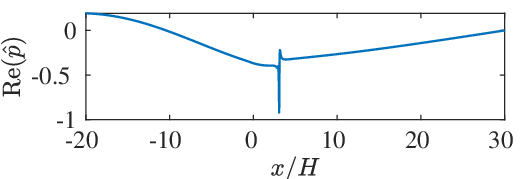}
        \put(2,35){(\textit{c},1)}
        \end{overpic}  
        \phantomsubcaption 
    \end{subfigure}
    \hspace{0.2pt}
    \begin{subfigure}[b]{0.42\textwidth}
        \centering
        \begin{overpic}[width=1\textwidth]{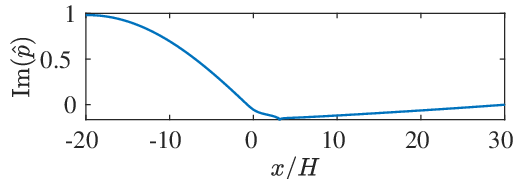}
        \put(2,35){(\textit{c},2)}
        \end{overpic}  
        \phantomsubcaption 
    \end{subfigure}    
    \vspace{0.1pt} 
    \begin{subfigure}[b]{0.42\textwidth}
        \centering
        \begin{overpic}[width=1\textwidth]{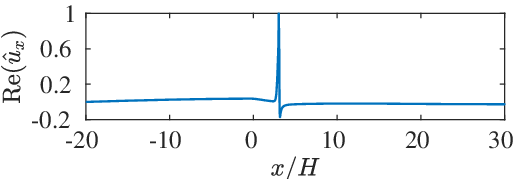}
        \put(2,35){(\textit{d},1)}
        \end{overpic}  
        \phantomsubcaption 
    \end{subfigure}
    \hspace{0.2pt} 
    \begin{subfigure}[b]{0.42\textwidth}
        \centering
        \begin{overpic}[width=1\textwidth]{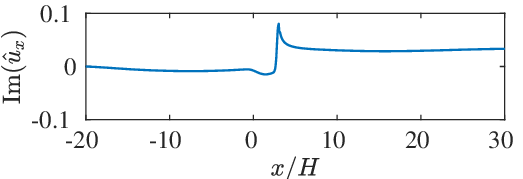}
        \put(2,35){(\textit{d},2)}
        \end{overpic}  
        \phantomsubcaption 
    \end{subfigure}
\caption{Pressure and streamwise velocity profiles of dominant unstable thermoacoustic mode in cases L50M2 (\textit{a},~\textit{b}) and L50M3 (\textit{c},~\textit{d}). Pressure on the central line (\textit{a},~\textit{c}). Streamwise velocity on the central line (\textit{b},~\textit{d}). Left and right columns correspond to the real and imaginary parts, respectively. All results are normalized by its corresponding maximum of $\left| \hat{p} \right| $ or $\left| \hat{u}_{x} \right| $ on the central line.}
\label{fig:Global_TA_l50_Ma0.02}
\end{figure}

\begin{figure}
    \centering
    \begin{subfigure}[b]{0.82\textwidth}
        \begin{overpic}[width=1\textwidth]{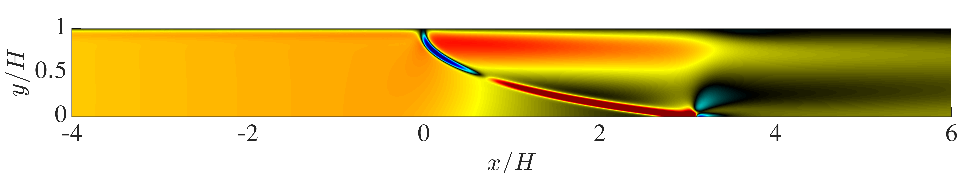}
        \put(2,17){(\textit{a})}
        \end{overpic}    
        \vspace{0pt}
        \begin{overpic}[width=1\textwidth]{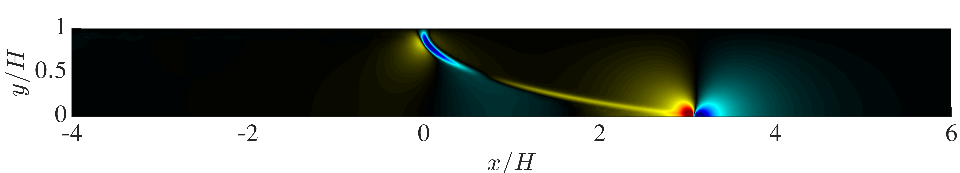}
        \put(2,17){(\textit{b})}
        \end{overpic}
        \vspace{0pt}
        \begin{overpic}[width=1\textwidth]{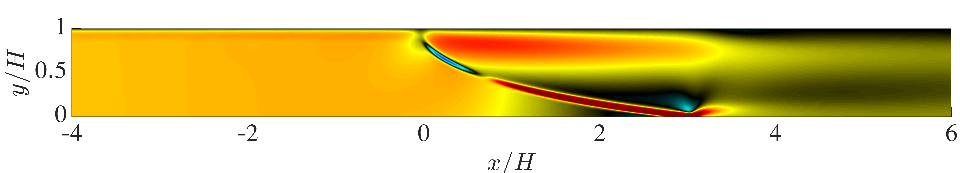}
        \put(2,17){(\textit{c})}
        \end{overpic}        
        \phantomsubcaption        
    \end{subfigure}%
    \hspace{5pt}                       
    \begin{subfigure}[b]{0.1\textwidth}
\includegraphics[width=\textwidth,height=0.32\textheight]{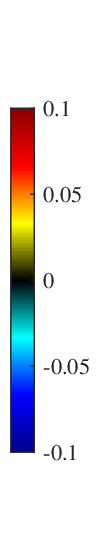}
    \phantomsubcaption
    \end{subfigure}  
\caption{Real part of streamwise velocity field and Helmholtz decomposition of dominant unstable thermoacoustic mode with fully reflective boundary conditions in case L50M2. Streamwise velocity field $\hat{u}_{x}$ (\textit{a}), potential component $\hat{u}_{d,x}$ (\textit{b}) and solenoidal component $\hat{u}_{s,x}$ (\textit{c}). All results are normalized by the maximum of $\left| \hat{u}_{x} \right| $.}
\label{fig:TA_reflective_l50_Ma0.02_Helm}
\end{figure}

\begin{figure}
    \centering
    \begin{subfigure}[b]{0.82\textwidth}
        \begin{overpic}[width=1\textwidth]{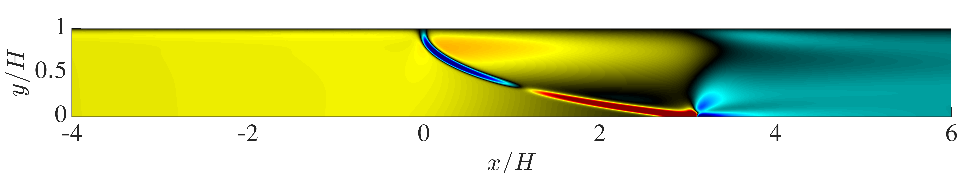}
        \put(2,17){(\textit{a})}
        \end{overpic}    
        \vspace{0pt}
        \begin{overpic}[width=1\textwidth]{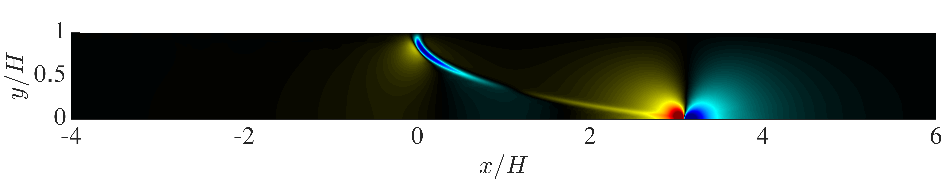}
        \put(2,17){(\textit{b})}
        \end{overpic}
        \vspace{0pt}
        \begin{overpic}[width=1\textwidth]{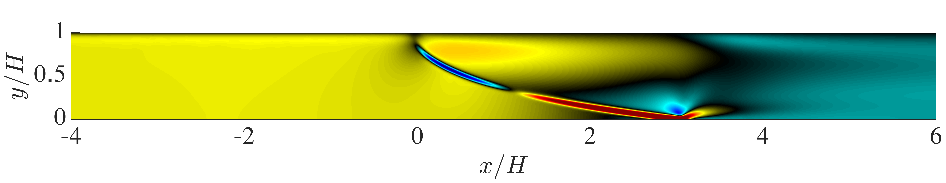}
        \put(2,17){(\textit{c})}
        \end{overpic}        
        \phantomsubcaption        
    \end{subfigure}%
    \hspace{5pt}                       
    \begin{subfigure}[b]{0.1\textwidth}
\includegraphics[width=\textwidth,height=0.32\textheight]{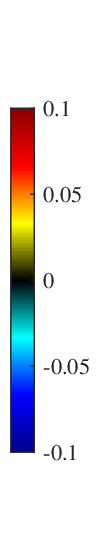}
    \phantomsubcaption
    \end{subfigure}  
  \caption{Real part of streamwise velocity field and Helmholtz decomposition of dominant unstable thermoacoustic mode with fully reflective boundary conditions in case L50M3. Streamwise velocity field $\hat{u}_{x}$ (\textit{a}), potential component $\hat{u}_{d,x}$ (\textit{b}) and solenoidal component $\hat{u}_{s,x}$ (\textit{c}). All results are normalized by the maximum of $\left| \hat{u}_{x} \right| $.}
\label{fig:TA_reflective_l50_Ma0.03_Helm}
\end{figure}

Figure~\ref{fig:Global_TA_l50_Ma0.02} presents the global mode profiles of pressure and streamwise velocity along the centerline of the acoustic duct. It is notable that the global velocity mode exhibits a highly compact structure near the flame, devoid of harmonic behavior within the duct, a characteristic consistent with that of an ITA mode, as depicted in figure~\ref{fig:Global_ITA_l50_Ma0.01}. This observation qualifies the unstable mode in case L50M2 probably as one originating from an ITA mode. Further analysis through Helmholtz decomposition of the near-flame velocity fields, presented in figure~\ref{fig:TA_reflective_l50_Ma0.02_Helm}, reinforces this identification, revealing decomposed fields that share the key attributes of a typical ITA mode, as seen in figure~\ref{fig:ITA_reflective_l50_Ma0.01_Helm}, especially the potential part. Interestingly, this unstable thermoacoustic mode also incorporates features characteristic of an AC mode. The potential component, $\hat{u}_{d,x}$, not only displays the notable dipole structure at the flame tip but also exhibits significant perturbation magnitude around the flame base. Additionally, the solenoidal component, $\hat{u}_{s,x}$, shows a transverse wave in the post-flame region, likely originating from an AC mode (see figure~\ref{fig:ac_reflective_l50_Ma0.01_Helm}), superimposed on the shape of ITA mode. This mode superposition phenomenon is attributed to the fact that this mode arises from a trajectory in close proximity to the exceptional point. Previous studies~\citep{orchini2020thermoacoustic,silva2023intrinsic} have suggested that thermoacoustic modes associated with exceptional points may share characteristics of both types of eigenmodes. Our detailed analysis of the near flame mode shapes, grounded in the linear analysis of reactive flows, substantiates this hypothesis. Figure~\ref{fig:TA_reflective_l50_Ma0.03_Helm} presents the decomposed fields of the unstable thermoacoustic mode in case L50M3. The results reveal similar flow patterns to those in figure~\ref{fig:TA_reflective_l50_Ma0.02_Helm}, with the structures associated with the ITA mode being more pronounced, due to the mode arising from a trajectory further from the exceptional point. This finding suggests that, as the locus of a thermoacoustic mode moves away from the exceptional point, the eigenmode predominantly exhibits characteristics of a single type, either ITA or AC mode. It also implies that the interplay between modes of different origins can be effectively captured in the near flame mode structures, even near the exceptional point. Therefore, we could analyze the structure of eigenmodes with Helmholtz decomposition, which serves as a robust and effective way to help us categorize mode origins and investigate mode interplay near exceptional points.

\section{Conclusions}
\label{Conclusions}
In this study, we employ a biglobal stability analysis framework based on the compressible reactive flow equations to investigate a ducted premixed flame, focusing on the ITA mode and the interplay between the ITA and AC modes in the vicinity of an exceptional point. We begin by identifying the pure ITA mode using the $-\pi$ criterion in a fully anechoic configuration. The mode's detailed features are subsequently extracted and characterized using Helmholtz decomposition. The decomposition reveals that the ITA mode is primarily dominated by its solenoidal component, while its potential component is mainly concentrated at the flame tip, exhibiting two distinct out-of-phase acoustic sources. Importantly, a parametric study demonstrates that the characteristic structures of the ITA mode are preserved despite changes in the acoustic reflectivity of boundary conditions, as well as increases in duct length and inflow Mach number. This finding suggests that these structures could serve as robust indicators for detecting thermoacoustic modes originating from the ITA mode.

The interaction between ITA and AC modes is further examined by varying parameters such as the inflow Mach number and the acoustic reflection coefficients at the boundaries. Strong mode veering and abrupt mode switching behaviors are observed near the exceptional points. However, depending on the specific path of parameter variation, the identified origin of the unstable thermoacoustic mode can vary, even when the same parameter is used. This inconsistency is particularly pronounced under critical parameter values corresponding to the exceptional point, due to its associated infinite sensitivity.

Given the unreliability of identifying thermoacoustic mode origins based solely on parameter variation, we examine the structures of unstable eigenmodes using decomposed flow fields under different operating conditions. It is evident that thermoacoustic modes originating from the AC mode possess flow fields that are distinct from those originating from the ITA mode. For the AC mode, the potential component displays strong velocity perturbations at the flame tip, while the solenoidal component is characterized by the development of shear layers interacting with non-planar acoustic waves. Additionally, the AC mode typically exhibits large scale perturbations to the natural acoustic mode throughout the entire duct, in contrast to the ITA mode, which shows highly localized velocity perturbations confined near the flame front. Leveraging these identified mode features, we analyze the unstable thermoacoustic mode near the exceptional point and find that the eigenmode shape incorporates characteristics of both ITA and AC modes. This indicates that the exceptional point affects not only the trajectories of the eigenmodes but also their shapes.

A significant contribution of this work lies in demonstrating the effectiveness of stability analysis using compressible reactive Navier-Stokes equations for studying the behavior of ITA and AC modes, as well as their interaction near the exceptional point. This approach allows for the detailed characterization of mode structures and features associated with thermoacoustic modes originating from various mechanisms. Notably, the distinct features revealed through flow field decomposition could be used to categorize the nature of these thermoacoustic modes, which addresses the challenge of parameter variation in mode origin identification. Additionally, some mode features identified through compressible reactive flow analysis may be integrated into low-order thermoacoustic models, thereby enhancing mode origin identification and improving the predictive accuracy of thermoacoustic instability.

While the Helmholtz decomposition is employed for mode feature extraction, it is important to note that this method is a purely kinematic decomposition applied to velocity fields and does not differentiate between different sources of compressible effects. In practical thermoacoustic systems, acoustic and hydrodynamic perturbations coexist with entropy and species perturbations related to flame dynamics. Recent studies~\citep{brokof2023towards}, suggest that momentum potential theory may offer an alternative approach to address the limitations of Helmholtz decomposition.

\backsection[Acknowledgements]{L.C acknowledges helpful discussions with Ming Yu from Tsinghua University and Ting Wu and Guowei He from Institute of Mechanics, Chinese Academy of Sciences on the Helmholtz decomposition and appropriate treatments regarding the boundary conditions. We acknowledge the suggestions and help from the FreeFEM Community on the implementation of parallel computation with FreeFem++. }

\backsection[Funding]{The funding supports by  the International Partnership Program of Chinese Academy of Sciences (No. 025GJHZ2022112FN) and the Strategic Priority Research Program of Chinese Academy of Sciences (No. XDA0380602) are acknowledged.}

\backsection[Declaration of interests]{The authors report no conflict of interest.}

%\backsection[Author ORCIDs]{Authors may include the ORCID identifers as follows.  F. Smith, https://orcid.org/0000-0001-2345-6789; B. Jones, https://orcid.org/0000-0009-8765-4321}

%\backsection[Author contributions]{Authors may include details of the contributions made by each author to the manuscript}

\appendix
\section{Boundary conditions for Helmholtz decomposition}\label{appA}
The boundary conditions for Helmholtz decomposition are discussed first in this appendix. Two Poisson equations should be solved here to conduct the Helmholtz decomposition:
\begin{equation}
\bnabla^{2}\hat{\varphi} = \bnabla \cdot \hat{\boldsymbol{u}} , \quad \bnabla^{2} \hat{\boldsymbol{\psi}} = -\bnabla \times \hat{\boldsymbol{u}}  \;.
\end{equation}
Then the decomposed velocity fields can be obtained by $\hat{\boldsymbol{u}}_{d} = \bnabla \hat{\varphi}$ and $\hat{\boldsymbol{u}}_{s} = \bnabla \times \hat{\boldsymbol{\psi}}$. The finite elements method is employed to solve the Poisson equations and the test functions are denoted as $\hat{v}_{\varphi}$ and $\hat{v}_{\psi}$. The weak forms are derived with integral by parts:
\begin{equation}
\int \boldsymbol{n} \cdot \bnabla \hat{\varphi} \ {\rm d}l - \int \bnabla \hat{\varphi} \cdot \bnabla \hat{v}_{\varphi} \ {\rm d}s = \int (\bnabla \cdot \hat{\boldsymbol{u}}) \hat{v}_{\varphi} \ {\rm d}s \;, 
\end{equation}
\begin{equation}
\int \boldsymbol{n} \cdot \bnabla \hat{\psi}_{z} \ {\rm d}l - \int \bnabla \hat{\psi}_{z} \cdot \bnabla \hat{v}_{\psi} \ {\rm d}s = \int -\hat{\omega}_{z} \hat{v}_{\psi} \ {\rm d}s \;.
\end{equation}
We only consider the two dimensional configuration in the present study. $\boldsymbol{n}$ denotes the normal vector, $\hat{\psi}_{z}$ refers to the spanwise component of the vector potential and $\hat{\omega}_{z}$ is the spanwise vorticity. To impose appropriate boundary conditions to solve the Poisson equations, we want to ensure both $\hat{\boldsymbol{u}}_{d}$ and $\hat{\boldsymbol{u}}_{s}$ vanish at the wall. The natural boundary conditions $\boldsymbol{n} \cdot \bnabla \hat{\varphi} \big|_{wall} = 0$ and $\boldsymbol{n} \cdot \bnabla \hat{\psi}_{z} \big|_{wall} = 0$ 
at the wall only constrains that $\hat{u}_{d,y}\big|_{wall} = 0 $ and $\hat{u}_{s,x}\big|_{wall} = 0$. As a result, we will not directly use $\bnabla \hat{\varphi}$ to calculate $\hat{u}_{d,x}$, because the streamwise velocity of potential part $\hat{u}_{d,x}$ may not be zero at the wall with this method. Alternatively, it will be evaluated by $\hat{u}_{d,x} = \hat{u}_{x} - \hat{u}_{s,x}$. According to the fact that $\hat{u} \big|_{wall} = 0$ and $\hat{u}_{s,x}\big|_{wall} = 0$, it could be shown that $\hat{u}_{d,x} \big|_{wall} = (\hat{u} - \hat{u} _{s,x}) \big|_{wall} = 0$ and the streamwise velocity of potential part vanishes at the wall. The normal velocity of solenoidal part $\hat{u}_{s,y}$ could be obtained with similar method $\hat{u}_{s,y}=\hat{u}_{y}-\hat{u}_{d,y}$.  

\section{Influence of entropy to acoustic reflection on the outlet boundary}\label{appB}
\begin{figure}
  \centerline{\includegraphics[width=0.6\textwidth]{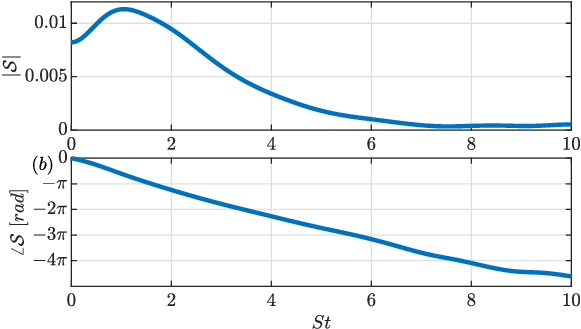}}
  \caption{Entropy transfer function determined from the linearized input-output analysis.
  (\textit{a}) Gain and (\textit{b}) Phase with the blue solid lines.}
\label{fig:ETF}
\end{figure}

\begin{figure}
  \centerline{\includegraphics{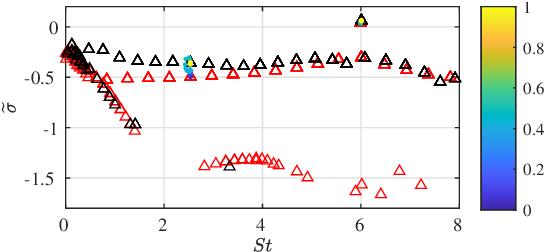}}
  \caption{Trajectories of thermoacoustic modes
   born of ITA mode and AC mode in case L50M1 and $R_{in}=1,R_{out}=-1$ with $R_{s}$ varying from 0 to 1. Eigenspectrums with different boundary conditions are marked by: $R_{s}=0$ (red $\triangle$ ) and $R_{s}=1$ (Black $\triangle$ ). The color of the trajectories corresponds to the value of $R_{s}$ during the parameter variation.  }
\label{fig:trajectory_Rs_l50_Ma0.01}
\end{figure}

In this appendix, we assess the influence of entropy waves, which may affect the thermoacoustic eigenspectra and mode trajectories. Entropy waves are generated at the perturbed flame front and can lead to acoustic reflections at the outlet boundary. The role of entropy waves in thermoacoustic instability has been extensively discussed in the literature~\citep{morgans2016entropy}. In our analysis, we have implicitly assumed that the acoustic wave reflection induced by entropy waves at the outlet boundary is minimal for the ducted flame configuration. Here, we provide a justification for this assumption. To this end, we first evaluate the entropy transfer function,
\begin{equation}
\mathcal{S}(\omega) = \frac{\hat{s}_{out}(\omega)/C_{p}}{\hat{u}_{x}(\omega,\boldsymbol{x}_{\mathrm{ref}})/\overline{u}_{x}(\boldsymbol{x}_{\mathrm{ref}})} \;,
\end{equation}
in which $\hat{s}_{out}$ denotes the entropy wave measured at the outlet of the acoustic duct. Here only perturbations of thermodynamic variables are considered:
\begin{equation}
\hat{s} = C_{p}\frac{\hat{T}}{\overline{T}} - R_{g} \frac{\hat{p}}{\overline{p}} \;,
\end{equation}
in which $R_{g}$ refers to the specific gas constant and the perturbations of species are ignored in our analysis. 
The entropy transfer function for case L10M1 is shown in figure~\ref{fig:ETF}.
The results are similar to the previous study on entropy generation in a premixed flame~\citep{meindl2021spurious}. It is observed that the amplitude of entropy wave at the outlet boundary remains relatively small, even when the length of the acoustic duct is short, provided that the dispersion and dissipation of the generated entropy wave are not significant. For a longer acoustic duct, the magnitude of the entropy wave at the outlet boundary is expected to be substantially smaller and its influence might be neglected in the thermoacoustic analysis of this configuration. Then we consider the case L50M1 to investigate the influence of entropy to acoustic reflection on the stability of the acoustic duct. Here the entropy to acoustic reflection coefficient $R_{s}$ is introduced and the boundary condition for the outlet boundary is replaced as:
\begin{equation}
\hat{g}_{out}-R_{out}\hat{f}_{out}- R_{s}\frac{\gamma \overline{p} \hat{s}_{out}}{C_{p}}=0 \;.
\end{equation}
The eigenspectrum for $R_{in}=1$ and $R_{out}=-1$ with different values of $R_{s}$ are shown in figure~\ref{fig:trajectory_Rs_l50_Ma0.01}. It is evident that the influence of $R_{s}$ is minor and can be reasonably ignored for this closed-open combustor configuration considered in the present study. However, it is worthy to mention that the influence of entropy wave could be significant in other combustor systems~\citep{morgans2016entropy}.

\section{Comparison with acoustic network models}\label{appC}

In this appendix, we present a comparison between the results obtained from the compressible reactive flow equations and those derived from acoustic network models. Given that the influence of entropy waves has been verified to be negligible, the Helmholtz equations are employed to construct the acoustic network models. An artificial coupling parameter, denoted as $\beta$ (referred to as $\mu$ in previous studies by \citep{emmert2017acoustic, ghani2021exceptional}), is introduced to the acoustic network models. For $\beta=0$, the system solutions reduce to a combination of the independent ITA and AC modes. In contrast, when $\beta=1$, the system corresponds to the desired thermoacoustic mode. For further details, interested readers are directed to the cited literature. The influence of the Mach number on the base flow is neglected, with the Mach number used to scale the eigenvalues of the acoustic network models as follows: 
\begin{equation}
St=\frac{\varpi H }{u_{0}}=\frac{\widetilde{\varpi} } {Ma \times (\frac{L}{H})} \;,
\end{equation}
in which $\alpha = \varsigma + i\varpi$ is the eigenvalue of acoustic network models and it is  non-dimensionalized as $\widetilde{\alpha} = \alpha L/c_{0}$. Other parameters required in this analysis are kept consistent with those in this study. The temperature ratio across the flame is $\theta = 5.54 $ and the ratio of specific acoustic impedance is $\xi = 2.55 $. The flame is located at $x_{flame}/L=0$ with the positions of inlet and outlet boundary as $x_{in}/L=-0.4$ and $x_{out}/L=0.6$. Here the length of acoustic duct is long enough to invoke the  compact flame assumption. The flame transfer function is fitted based on the measured samples at real frequencies shown in figure \ref{fig:FTF}, which is similar to the technique used in previous studies~\citep{hoeijmakers2014intrinsic,hoeijmakers2016flame}. Here we employ a $n-\tau-\varrho$ model and the fitted flame transfer function is expressed as: $\mathcal{F}(\varpi) = n \mathrm{exp}(-i\varpi\tau)\mathrm{exp}(-0.5\varpi^{2}\varrho^{2})$, with $n=1.0$, $\tau=1.25$ and $\varrho=0.3$. Here $\varrho$ represents the generalized Gaussian distribution. This model was first introduced to model the transfer function of a premixed swirl burner~\citep{alemela2010determination} and discussed in more details by~\citet{polifke2020modeling}. The system matrix $\mathcal{M}$ (the same as~\citep{ghani2021exceptional}) for the acoustic network models is written as:
\begin{equation}
\mathcal{M} = \left[ \begin{array}{cccc:cc}
-1 & R_{u} & 0 & 0 & 0 & \beta R_{u}\dfrac{\theta}{1+\xi} \\ 
\dfrac{1-\xi}{1+\xi} & -1 & 0 & \dfrac{2}{1+\xi} & 0 & 0  \vspace{4pt} \\ 
\dfrac{2\xi}{1+\xi} & 0 & -1 & \dfrac{\xi-1}{1+\xi} & 0 & \beta\dfrac{\xi\theta}{1+\xi} \vspace{4pt} \\ 
0 & 0 & R_{d} & -1 & 0 & 0  \vspace{4pt} \\
\hdashline
\beta & -\beta & 0 & 0 & -1 & -\dfrac{\theta}{1+\xi}  \vspace{4pt} \\
0 & 0 & 0 & 0 & \mathcal{F}(\alpha) & -1 
\end{array} \right ] \;,
\end{equation}
in which $R_{u}$ and $R_{d}$ refer to the reflection coefficients upstream and downstream of the flame, respectively. $R_{u}$ is related to reflection coefficient $R_{in}$ by $R_{u} = R_{in} \mathrm{exp}(-2\alpha (x_{flame}-x_{in})/c_{u})$, while $R_{d}$ is related to $R_{out}$ by $R_{d} = R_{out} \mathrm{exp}(-2\alpha(x_{out}-x_{flame})/c_{d})$. $c_{u}$ and $c_{d}$ denotes the sound speed upstream and downstream of the flame, respectively. The nonlinear eigenvalue problem can be expressed with this system matrix as follows:
\begin{equation}
(\mathcal{M}-\alpha \boldsymbol{I})\hat{\boldsymbol{q}}_{N} = 0 \;,
\end{equation}
in which $\hat{\boldsymbol{q}}_{N}$ is the corresponding eigenvector of the acoustic network models. This nonlinear eigenvalue problem is then solved with a contour integral method~\citep{buschmann2020solution} with a numerical tolerance on the magnitude of the determinant of the matrix $(\mathcal{M}-\alpha \boldsymbol{I})$ set to less than $10^{-6}$. 

\begin{figure}
  \centerline{\includegraphics[width=0.7\textwidth]{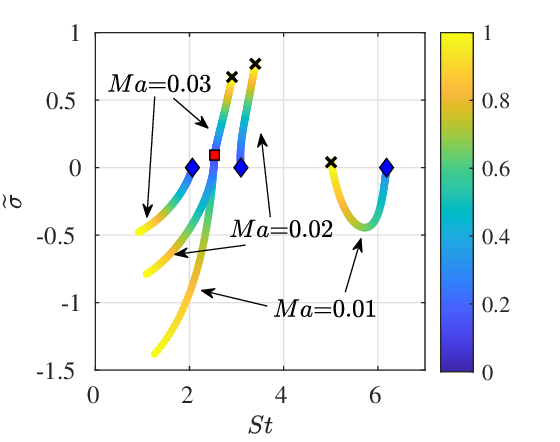}}
  \caption{The results of acoustic network models for trajectories of ITA mode and AC modes during parameter variation for $\beta$ from 0 to 1 in case L50M1, L50M2 and L50M3. The pure ITA modes and pure AC modes at $\beta = 0$ are marked by red $\square$ and blue $\Diamond$. The unstable thermoacoustic modes are marked by black $\times$. The color of the trajectories corresponds to the value of $\beta$ during the parameter variation.}
\label{fig:Acoustic_Network_miu_continuation}
\end{figure}

\begin{figure}
  \centerline{\includegraphics[width=0.7\textwidth]{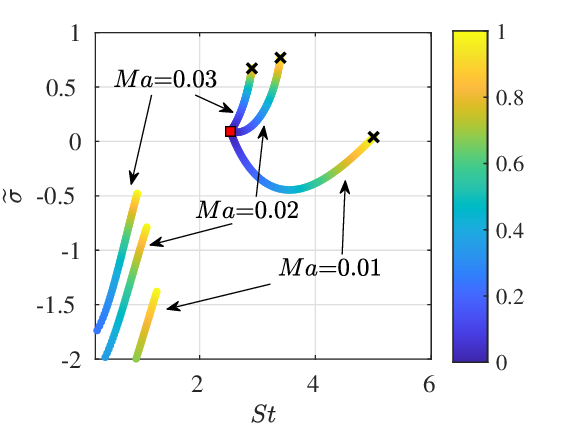}}
  \caption{The results of acoustic network models for trajectories of ITA mode and AC modes during parameter variation for $R_{k}$ from 0 to 1 in case L50M1, L50M2 and L50M3. The pure ITA modes are marked by red $\square$. The unstable thermoacoustic modes are marked by black $\times$. The color of the trajectories corresponds to the value of $R_{k}$ during the parameter variation.}
\label{fig:Acoustic_Network_Rk_continuation}
\end{figure}

The results of parameter variations with respect to  $\beta$ and $R_k$ are presented in figure~\ref{fig:Acoustic_Network_miu_continuation} and~\ref{fig:Acoustic_Network_Rk_continuation}. In general, the acoustic network model captures the loci of unstable thermoacoustic modes in the eigenspectrum. However, different parameter choices lead to varying conclusions regarding mode origin. When $\beta$ is varied, only the unstable thermoacoustic mode in case L50M3 is found to originate from the ITA mode, while the unstable thermoacoustic modes in cases L50M1 and L50M2 are both arsing from AC modes. In contrast, varying $R_k$ reveals that all unstable thermoacoustic modes in the three cases arise from the ITA mode. This inconsistency in the identification of thermoacoustic modes under parameter variation underscores the unreliability of using parameter variation alone to classify mode origins, particularly within the framework of acoustic network models, which neglect important aspects of flame dynamics and hydrodynamic flows, especially in parameter variation with reflection coefficients~\citep{silva2023intrinsic}. Another noteworthy feature is the veering of the AC mode trajectory in case L50M1 at $St \approx 5.5$. This behavior, previously unobserved, can be attributed to the presence of the second ITA mode and its interaction with the AC mode, similar to the mode interplay observed in the earlier study~\citep{sogaro2019thermoacoustic}. The eigenfrequency of the second ITA mode coincides with the frequency where the phase of $\mathcal{F}(\varpi)$ equals to $-3\pi$. While this eigenmode is damped in the biglobal analysis of the compressible flow equations, it persists in the solutions of the acoustic network models. Therefore, we choose not to depict the trajectory of the second ITA mode and focus instead on the interaction between the first ITA and AC modes.

\begin{figure}
    \centering
    \begin{subfigure}[b]{0.42\textwidth}
        \centering
        \begin{overpic}[width=1\textwidth]{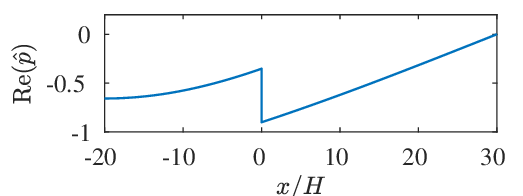}
        \put(2,35){(\textit{a})}
        \end{overpic}  
        \phantomsubcaption 
    \end{subfigure}
    \hspace{0.2pt}
    \begin{subfigure}[b]{0.42\textwidth}
        \centering
        \begin{overpic}[width=1\textwidth]{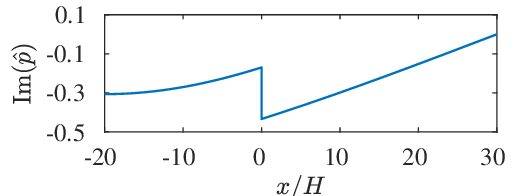}
        \put(2,35){(\textit{b})}
        \end{overpic}  
        \phantomsubcaption 
    \end{subfigure}    
    \vspace{0.1pt} 
    \begin{subfigure}[b]{0.42\textwidth}
        \centering
        \begin{overpic}[width=1\textwidth]{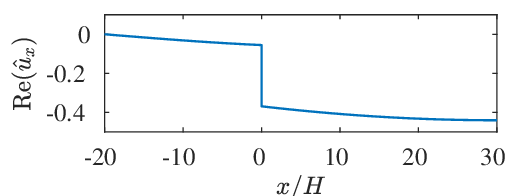}
        \put(2,35){(\textit{c})}
        \end{overpic}  
        \phantomsubcaption 
    \end{subfigure}
    \hspace{0.2pt} 
    \begin{subfigure}[b]{0.42\textwidth}
        \centering
        \begin{overpic}[width=1\textwidth]{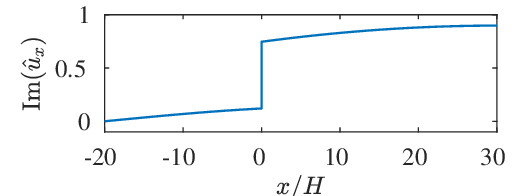}
        \put(2,35){(\textit{d})}
        \end{overpic}  
        \phantomsubcaption 
    \end{subfigure}
\caption{The results of pressure and streamwise velocity profiles of the thermoacoustic mode originating from AC mode in case L50M1 obtained by acoustic network models. Pressure (\textit{a},\textit{b}) and streamwise velocity (\textit{c},\textit{d}). Real part (\textit{a},\textit{c}) and imaginary part (\textit{b},\textit{d}). All results are normalized by its corresponding maximum of $\left| \hat{p} \right| $ or $\left| \hat{u}_{x} \right| $.}
\label{fig:Acoustic_Network_ac_l50_Ma0.01}
\end{figure}

\begin{figure}
    \centering
    \begin{subfigure}[b]{0.42\textwidth}
        \centering
        \begin{overpic}[width=1\textwidth]{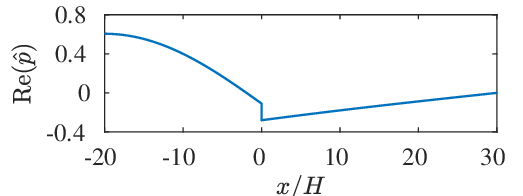}
        \put(0,38){(\textit{a},1)}
        \end{overpic}  
        \phantomsubcaption 
    \end{subfigure}
    \hspace{0.2pt}
    \begin{subfigure}[b]{0.42\textwidth}
        \centering
        \begin{overpic}[width=1\textwidth]{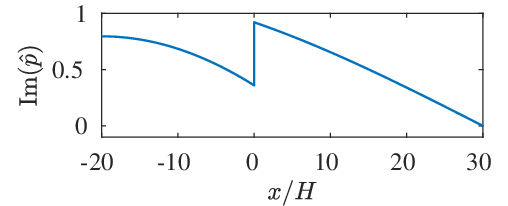}
        \put(0,38){(\textit{a},2)}
        \end{overpic}  
        \phantomsubcaption 
    \end{subfigure}    
    \vspace{0.1pt} 
    \begin{subfigure}[b]{0.42\textwidth}
        \centering
        \begin{overpic}[width=1\textwidth]{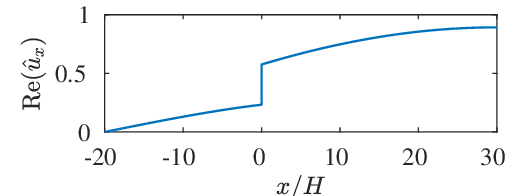}
        \put(0,38){(\textit{b},1)}
        \end{overpic}  
        \phantomsubcaption 
    \end{subfigure}
    \hspace{0.2pt} 
    \begin{subfigure}[b]{0.42\textwidth}
        \centering
        \begin{overpic}[width=1\textwidth]{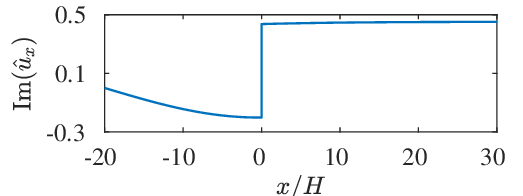}
        \put(0,38){(\textit{b},2)}
        \end{overpic}  
        \phantomsubcaption 
    \end{subfigure}
    \begin{subfigure}[b]{0.42\textwidth}
        \centering
        \begin{overpic}[width=1\textwidth]{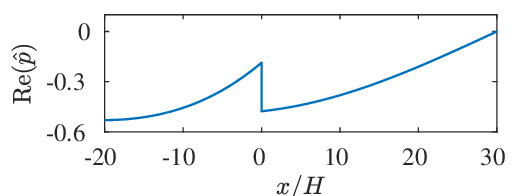}
        \put(0,38){(\textit{c},1)}
        \end{overpic}  
        \phantomsubcaption 
    \end{subfigure}
    \hspace{0.2pt}
    \begin{subfigure}[b]{0.42\textwidth}
        \centering
        \begin{overpic}[width=1\textwidth]{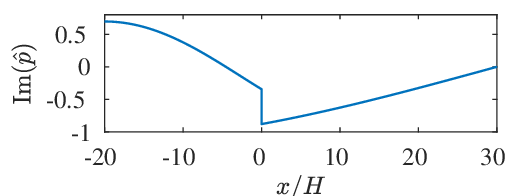}
        \put(0,38){(\textit{c},2)}
        \end{overpic}  
        \phantomsubcaption 
    \end{subfigure}    
    \vspace{0.1pt} 
    \begin{subfigure}[b]{0.42\textwidth}
        \centering
        \begin{overpic}[width=1\textwidth]{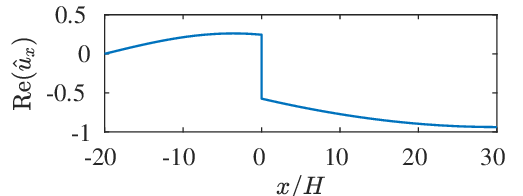}
        \put(0,38){(\textit{d},1)}
        \end{overpic}  
        \phantomsubcaption 
    \end{subfigure}
    \hspace{0.2pt} 
    \begin{subfigure}[b]{0.42\textwidth}
        \centering
        \begin{overpic}[width=1\textwidth]{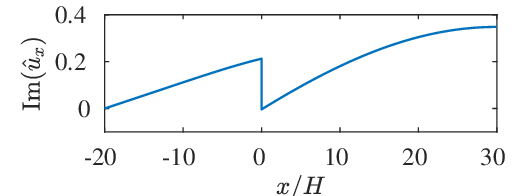}
        \put(0,38){(\textit{d},2)}
        \end{overpic}  
        \phantomsubcaption 
    \end{subfigure}
\caption{Pressure and streamwise velocity profiles of dominant unstable thermoacoustic mode obtained by acoustic network models in cases L50M2 (\textit{a},~\textit{b}) and L50M3 (\textit{c},~\textit{d}). Pressure (\textit{a},~\textit{c}) and streamwise velocity (\textit{b},~\textit{d}). Left and right columns correspond to the real and imaginary parts, respectively. All results are normalized by its corresponding maximum of $\left| \hat{p} \right| $ or $\left| \hat{u}_{x} \right| $ on the central line.}
\label{fig:Acoustic_Network_ac_l50_Ma0.02}
\end{figure}

We continue to investigate the mode shapes of unstable thermoacoustic modes to determine whether specific features can be used to distinguish between thermoacoustic modes originating from ITA and AC modes. The pressure and streamwise velocity profiles of the corresponding unstable thermoacoustic mode, as predicted by the acoustic network models for case L50M1, are presented in figure~\ref{fig:Acoustic_Network_ac_l50_Ma0.01}. It is expected that the results obtained from the acoustic network models differ significantly from those of the compressible flow equations shown in figure~\ref{fig:Global_ac_l50_Ma0.01}, owing to several simplifying assumptions. Specifically, the acoustic waves and flame are approximated as one-dimensional, and all hydrodynamic effects are encapsulated within the flame transfer function in the acoustic network models. These approximations inevitably neglect some physical mechanisms, particularly the influence of hydrodynamic flows near the flame front, making the mode shapes derived from the acoustic network models less suitable for precise identification of thermoacoustic modes. Nevertheless, certain features can still be captured by the acoustic network models. The pressure and streamwise velocity profiles of the unstable thermoacoustic modes for cases L50M2 and L50M3 are shown in figure~\ref{fig:Acoustic_Network_ac_l50_Ma0.02} and compared with those in figure~\ref{fig:Global_TA_l50_Ma0.02}. For thermoacoustic modes originating from ITA modes, the streamwise velocity reverses direction across the flame in both L50M2 and L50M3, consistent with previous studies utilizing phasor diagrams to categorize thermoacoustic modes~\citep{yong2023categorization}. In contrast, in case L50M1, the streamwise velocity remains unidirectional, indicating that the corresponding unstable thermoacoustic mode originates from the AC mode. Hence, the eigenmode shapes are able to offer further insights into the origins of different thermoacoustic modes, even within the framework of simplified acoustic network models.

\section{Validation of the codes}\label{appD}
\begin{figure}
    \centering
    \begin{subfigure}{0.7\textwidth}
        \centering
        \centerline{
        \begin{overpic}[width=1\textwidth]{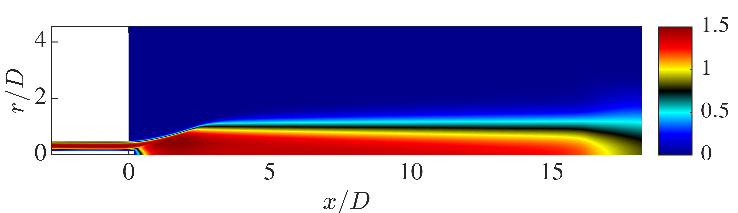}
        \put(2.5,26){(\textit{a})}
        \end{overpic}
        }
        \phantomsubcaption        
    \end{subfigure}
    \vspace{0pt}
    \begin{subfigure}{0.7\textwidth}
        \centering
        \centerline{
        \begin{overpic}[width=1\textwidth]{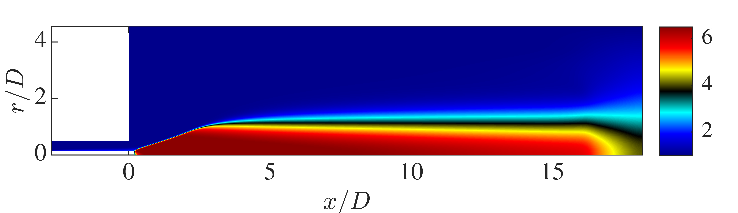}
        \put(2.5,26){(\textit{b})}
        \end{overpic}
        }
        \phantomsubcaption        
    \end{subfigure}   
\caption{Baseflow solutions of the V flame at $\Rey=2282$ obtained by our solver. (\textit{a}) Streamwise velocity and (\textit{b}) Temperature.}
\label{fig:Validation_baseflow}
\end{figure}

\begin{figure}
\centerline{{\includegraphics[width=0.7\textwidth]{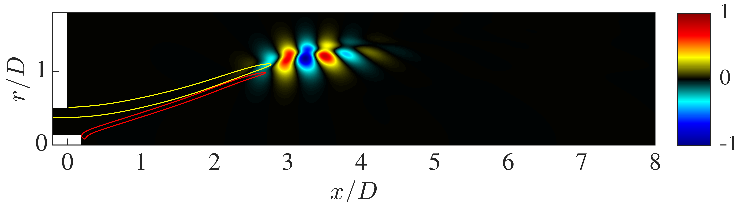}}}
    \caption{Real part of radial velocity fields of the leading flame-tip mode at $\Rey=2282$ obtained by our solver. All results are normalized by its corresponding maximum. The red contour indicates the position of shear layer and the yellow contour indicates the position of flame front, similar with the cited literature.}
\label{fig:Validation_flame_tip_mode}
\end{figure}

The code validations are extended to both the nonlinear solver for steady state flame calculations and the linear solver for the biglobal stability analysis, with an emphasis on validating the numerical implementations. For this purpose, a premixed annular V-flame at a Reynolds number of $Re = 2282$, as reported in the literature~\citep{wang2024onset}, is considered. The compressible Navier-Stokes equations in the low Mach number limit are solved. We employ the 1S$_{-}$CH4$_{-}$MP1 chemical mechanism, consistent with the referenced study, which thereby allows to validate the chemistry treatment in our solver as well. All parameters and settings are maintained identically to those in the previous study. The steady state solutions for streamwise velocity and temperature are obtained and presented in figure~\ref{fig:Validation_baseflow}. The leading flame-tip mode is captured, and the corresponding radial velocity profiles are shown in figure~\ref{fig:Validation_flame_tip_mode}. These results exhibit very good agreement with those reported by~\citet{wang2024onset} (see figures 3 and 6 of that work for comparison). Quantitatively, the accuracy is assessed by comparing the eigenvalue of a representative eigenmode. The eigenvalue of the leading flame-tip mode obtained using our solver is $\widetilde{\sigma}/(2\pi) = -0.0326$ and $\widetilde{St} = St/(2\pi) = 0.3307$. The relative error is below 5\% compared to the reported eigenvalue in~\citet{wang2024onset} (see figure 4 of that work for comparison). 

%\bibliographystyle{jfm}
%\bibliography{jfm}
%Use of the above commands will create a bibliography using the .bib file. Shown below is a bibliography built from individual items.

\bibliographystyle{jfm}
\bibliography{jfmLu}

%% End of file `jfm2esam.bib'.

\end{document}